\documentclass[twocolumn,pre,floats,aps,amsmath,amssymb,nofootinbib,usenatbib]{emulateapj}
\usepackage{graphicx}
\usepackage{bm}
\usepackage{natbib}
\usepackage{hyperref}
\usepackage[english]{babel}
\usepackage{tabu}
\usepackage{amsmath}
\usepackage{float}
\usepackage{color}
\usepackage{hyperref}

\begin{document}    
%%%%%%%%%%%%%%%%%%%%%%%%

%%%%%%% Arxiver %%%%
%@arxiver{Figure_6.png,Figure_3.png}
%%%%%%%%%%%%%%%

\title{Attenuation modified by DIG and dust as seen in M31}

\author{Neven Tomi\v{c}i\'{c}$^{1}$,
        Kathryn Kreckel$^{1}$,
        Brent Groves$^{2}$,
        Eva Schinnerer$^{1}$,
        Karin Sandstrom$^{3}$,
        Maria Kapala$^{4}$,
        Guillermo A. Blanc$^{5,6,7}$,        
        Adam Leroy$^{8}$,
        }

\affil{
$^1$Max Planck Institute for Astronomy (MPIA), K\"{o}nigstuhl 17, Heidelberg, Germany
\\$^2$Australian National University, Canberra, Australia
\\$^3$Center for Astrophysics and Space Sciences, University of California, USA
\\$^4$Department of Astronomy, University of Cape Town, Republic of South Africa
\\$^5$Observatories of the Carnegie Institution for Science, 813 Santa Barbara St, Pasadena, CA, 91101, USA.
\\$^6$Departamento de Astronom\'{i}a, Universidad de Chile, Camino del Observatorio 1515, Las Condes, Santiago, Chile
\\$^7$Centro de Astrof'sica y Tecnolog\'{i}as Afines (CATA), Camino del Observatorio 1515, Las Condes, Santiago, Chile
\\$^8$National Radio Astronomy Observatory, Charlottesville, USA
}
\email{tomicic@mpia-hd.mpg.de}

\date{\today}
\shorttitle{Attenuation modified by DIG and dust as seen in M31}
\shortauthors{Tomi\v{c}i\'{c} et al.}

\begin{abstract}

  The spatial distribution of dust in galaxies affects the global attenuation, and hence inferred properties, of galaxies. We trace the spatial distribution of dust in five fields (at 0.6-0.9 kpc scale) of M31 by comparing optical attenuation with the total dust mass distribution. We measure the attenuation from the Balmer decrement using Integral Field Spectroscopy and the dust mass from Herschel far-IR observations. Our results show that  M31's dust attenuation  closely follows a  foreground screen model, contrary to what was previously found in other nearby galaxies. By smoothing the M31 data we find that spatial resolution is not the cause for this difference. Based on the emission line ratios and two simple models, we conclude that previous models of dust/gas geometry need to include a weakly or non-attenuated diffuse ionized gas (DIG) component. Due to the variation of dust and DIG scale heights with galactic radius, we conclude that different locations in galaxies will have different vertical distributions of gas and dust and therefore different measured attenuation. The difference between our result in M31 with that found in other  nearby galaxies can be explained by our fields in  M31 lying at larger galactic radii than the previous studies that focused  on the centers of galaxies.

\end{abstract}

\maketitle

\section{Introduction}
\label{sec:intro}

The attenuation and reddening by dust can severely  impair our understanding of galaxies and the ISM environment. Dust preferentially absorbs ultra-violet (UV) and optical photons and re-emits this radiation in the infra-red (IR). The final impact of this reprocessing on the spectral energy distribution of a galaxy is dependent upon the properties of the dust and its spatial distribution relative to the stars and gas (\citealt{Witt92}, \citealt{Gordon03}, \citealt{Draine11}). To understand the observed light from the galaxies, correct models of  dust properties, distribution and the resulting effect on the spectra are needed.   
 
  Extinction is the result of absorption and scattering of the light by dust along a single line of sight. The effect is more pronounced on light at shorter wavelengths, resulting in an overall reddening of the light (usually denoted by the selective extinction E$ _{\rm B-V} $). With different dust/gas geometrical distribution, the effects on observed light would be different.  The combined effects of extinction and geometry is usually called ``attenuation''. \citet{Caplan86}, \citet{Witt92}, \citet{Calzetti94}, \citet{Gordon03} and \citet{DraineBT11}  derived various models for the relative dust/gas distributions and the corresponding effects on the stellar light and observed attenuation. Among these models, two show extreme scenarios, where in one the dust and gas are not mixed and another where they are.     
  
  The `foreground screen' model  assumes that the dust is distributed as a thin screen  between the stars and the observer. This model represents the `Extinction' case where all light is either absorbed or scattered out of the line of sight, and A$_{\rm V}$ correlates linearly with dust mass surface density. However, if the dust screen is on the far side behind the stars compared to the observer, that will result in no extinction. 

 The `mixed media' model assumes the stars and dust are uniformly distributed and mixed. In this distribution some stars suffer relatively less extinction than others (i.e. closer to the observer) altering the attenuation. Also in this distribution light from stars can be scattered into the line of sight of the observer, also altering the attenuation.   
 
 \citet{Hulst88} and \citet{CalzettiKinney96} observed ratios of various hydrogen lines in 14 nearby galaxies. They found that integrated $ \sim $kpc regions in these galaxies typically had attenuations suggesting a dust distribution between the screen and mixed models.  
 
 \citet{Liu13} investigated the dust attenuation of HII regions in M83 using the ratio of the Balmer and Paschen lines from \textit{Hubble Space Telescope}/WFC3 narrow-band imaging at $\sim 6$\, pc spatial resolution.   They found a diverse range of geometries, where the center of M83  has a dust distribution closer to the mixed model while the outer radii have HII regions with attenuation closer to  the screen model. When averaged to $ \geq $100-200\,pc spatial resolution, their data follow a foreground screen model.   
 
  Using optically thin tracers, one simple way of deducing the spatial distribution of dust is to observe the effect of extinction on the known ratio of optical Balmer lines (H$ \alpha $, H$ \beta $, H$ \gamma $, H$ \delta $) and then compare it with the extinction expected from the dust mass distribution.

While previous works of \citet{Hulst88}, \citet{CalzettiKinney96} and \citet{Liu13} used only attenuation based on optical and near-infrared (NIR) lines to determine the dust distribution, \citet{Kreckel13} (hereafter K13) used two independent measures of dust - optical attenuation and IR emission to measure the distribution. In K13, eight nearby galaxies were observed as a part of the KINGFISH\footnote{\textbf{K}ey \textbf{I}nsight on \textbf{N}earby \textbf{G}alaxies: A \textbf{F}ar-\textbf{I}nfrared \textbf{S}urvey with \textbf{H}erschel}  project (\citealt{Kennicutt11})  with optical integral field  spectroscopy (IFS) and far-IR observations (done with \textit{Herschel Space Observatory}; \citealt{Griffin10},  \citealt{Pilbratt10}). They distinguish features at physical scales of $ \sim $1\,kpc within the galaxy disks. K13  conclude that the distribution of dust and gas in these galaxies lies somewhere between the screen and mixed models. 
 
Following the methodology of K13, we observed five fields in the Andromeda galaxy (M31) to determine the spatial distribution of dust as compared to the ionized gas at high spatial resolution ($ \approx $100\,pc or 24.9$ '' $). Compared to K13, our higher  resolution gives us the opportunity to resolve star-forming complexes at $ \approx $100\,pc resolution and HII regions at the $ \approx $10\,pc  resolution. With this resolution,  we are able to trace different environments of  dust and ionising gas. We analyze the dust distribution using  two extreme dust models (foreground screen and mixed model) and compare our results with those from K13. 
   
 The Andromeda galaxy is the closest large spiral galaxy, providing the opportunity to observe  dust on physically smaller scales than those in K13. M31 is a massive ($\sim 10^{10.5}$\,M$_{\odot}$) SA(s)b galaxy with ring-like structures. Its distance from the Milky Way is $ \sim $780\,kpc \citep{Stanek98} and it is highly inclined ($\sim$70$^\circ$, \citealt{Dalcanton12}). R$ _{25} $ of M31 is $ \approx $20.5\,kpc \citep{Zurita12}. The total star-formation rate (SFR) for the entire disk of M31 is $ \sim 1$ M$ _{\odot} $yr$ ^{-1} $ (\citealt{Williams03},\citealt{Lewis15}).
  
 In this article, Section 2  presents an overview of the data and its reduction details. In Section 3 we show our main results and a comparison between the attenuation and dust mass column density. In Section 4 we discuss possible explanations of our results. The conclusions  are presented in Section 5, followed by a summary.

\section{Data} \label{Sec: Data}

The Andromeda Galaxy (M31) provides the best compromise between spatial resolution and a global view in the study of galaxy structure. The proximity of M31 to the Milky Way enables observation of its ISM with high resolution ($ \sim $10.2\,pc in the optical and $ \sim $100\,pc at 350$ \,\mu $m). 

To study the relationship between attenuation and dust in M31, we have combined optical integral field unit (IFU) spectroscopy with far-infrared imaging. The attenuation is traced using optical spectroscopy, and the dust mass surface density was independently derived from far-IR Spectral Energy Distribution (SED) fitting using \emph{Herschel} and \emph{Spitzer} (Spitzer Space Telescope) photometry.  

\begin{figure}[t]
\centering
\includegraphics[width=1.0\linewidth]{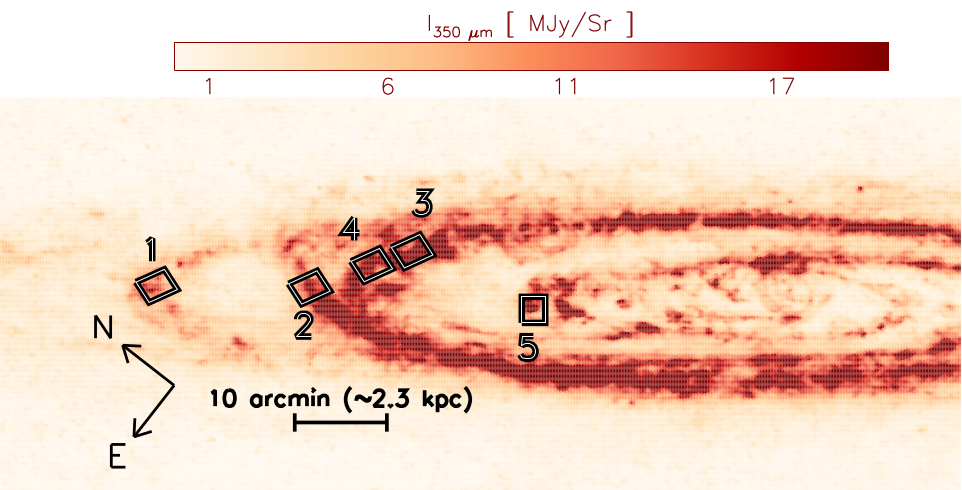}
\caption{Positions of the five fields in M31 used in this work, overlaid on a \textit{Herschel} SPIRE 350$ \,\mu m$ intensity map in 24.9$ '' $ resolution.\\}
\label{fig:FigM31}
\end{figure}

We targeted five  fields  in  M31, chosen to have a large suite of ancillary multiwavelength data (H$ \alpha $, 24$ \,\mu m$ and FUV) and to cover a  range of star formation rates and environments. 
The fields and data are also used by \citet{Kapala15}  to trace the origin of [CII] line emission, as part of the Survey of Lines in M31 (SLIM, PI Sandstrom K.). The  positions of the five fields are shown  in Fig. \ref{fig:FigM31} and Tab. \ref{tab:Tab01}. 

In the following subsections we describe the data reduction, flux calibration, and analysis of the spectra that are performed following the procedures outlined in K13 and \citet{Kapala15}.

\subsection{Optical Integral Field Spectroscopy}

 We observed all fields using the Potsdam Multi-Aperture Spectrophotometer (PMAS, \citealt{Roth05}) at the 3.5\,m telescope at the Calar Alto Observatory on September 16-24, 2011. To split the spatial image, the telescope uses a specialized fiber-bundle,  PPaK, which consists of 331 bare science fibers (with an additional 36 sky and 15 calibration fibers) in a hexagonal grid. The grid has a diameter of $75''$ \citep{Verheijen04} and  spatial sampling of  $2.7''$ per fiber.  We used the $ 4k\times4k $ CCD detector with the V300 grating to achieve a wavelength coverage of 3500-9000 \AA{} (centred at 5400 \AA{}) and resolution  of R$ =1000$.

 Each of the five observed fields combines 10 pointings in a mosaic, resulting in an area of $ 3'\times4' $ (680\,pc $ \times $ 900\,pc) for each field. Resulting mosaics have an effective PPaK resolution of 2.7$ '' $.  Each pointing was observed with a dither pattern (three dither positions shifted by $\rm \Delta Dec =+1.''56$, $\rm \Delta Dec =+0.''78$ and $\rm \Delta RA=+1.''56$, $\rm \Delta  RA=-0.''78$) with 2$ \times $600\,s exposures in order to fill in gaps between the fibers, thus covering the entire field of view. Dedicated sky observations (with 120 s exposures) were taken in the same manner between each science observation in order to be subtracted later during calibration. 
 
  Astrometry for each mosaic position was applied by eye through comparison of compact HII regions  with  H$ \alpha $ images from  the Local Group Galaxies Survey \citep{Azimlu11}. Additionally we compared \textit{r} and \textit{g} band SDSS images with our data by applying the SDSS response functions for the corresponding bands to our observed spectra.  Maximum deviation in astrometry are 2$ '' $, with a mean offset of around 1$ '' $. Astrometric inaccuracies at this level do not affect our data analysis as we compare the A$ _{\rm V} $ and dust maps at 25$ '' $ resolution.

 To translate the electron count values into fluxes, we observed the standard stars BD +33d2642 and BD +25d4655 \citep{Oke90}. 
 The positions of lines and spectra on the detector, the optical path and the transmission are all affected by fiber flexure of the IFU instrument \citep{Sanchez06}. To correct for these effects, we obtained calibration continuum lamp images (used for positioning of the spectra), He+HgCd arc lamp images (used for wavelengths calibration) and twilight flats (used for accurate flat fielding).

\begin{table}[t]
\centering
\caption{Coordinates and approximate distances\tablenotemark{a} from the galaxy center (in kpc) for our fields.  }
\begin{tabular}{cccc}
\hline 
    	     Field & R.A. & Dec. & R  \\ 
    	     &  (J2000) &  (J2000) & kpc \\ 
    	   \hline 1 &$ 00^{h}46^{m}28.88^{s}$ & $+42^{\circ}11'38.16''$ & 16   \\ 
    	          2 &$ 00^{h}45^{m}34.04^{s}$ & $+41^{\circ}58'33.53''$ & 12.2  \\ 
    	          3 &$ 00^{h}44^{m}36.04^{s}$ & $+41^{\circ}52'53.58''$ & 11.7  \\ 
    	          4 &$ 00^{h}44^{m}58.54^{s}$ & $+41^{\circ}55'09.14''$ & 11.8   \\ 
    	          5 &$ 00^{h}44^{m}25.58^{s}$ & $+41^{\circ}37'37.20''$ & 6.8   \\ 
    	   \hline 
    	   \\
\end{tabular}  \\
\label{tab:Tab01}

\tablenotetext{1}{R$ _{25} $ of M31 is $ \approx $20.5\,kpc \citep{Zurita12}.}
\end{table}

  The atmospheric conditions were mostly clear, resulting in approximately uniform imaging of our fields. The observations of some pointings were repeated on 24th of September due to bad weather (clouds) or moonlight contamination. Seeing was subfiber (less than $2.7''$) for all observations.

\subsubsection{Calibration }

 We reduce and calibrate all the data using the P3D software package\footnote{http://p3d.sourceforge.net/}, version 2.2.6.  \citep{Sandin10}.  For the first step we perform a bias correction using the median image of all bias frames as a master bias. Next, we obtain a master flat field from all twilight flat images. Observations are cleaned of cosmic rays following the L.A. Cosmic technique \citep{Dokkum01} as adapted within P3D. We verify that the cosmic ray removal algorithm implemented in P3D robustly cleans up the images and the corresponding noise maps. We then create the master trace mask to determine the position of all spectra on the CCD. The trace mask is constructed by stacking  multiple calibration continuum lamp images, and fitting the peaks of emission lines with Gaussian functions  along the cross-dispersion axis \citep{Sandin10}. To remove the possibility of overlapping Gaussians, a modified optimal extraction method is applied \citep{Horne86} that simultaneously fits all the line profiles. The He+HgCd arc lamp images are used for construction of the dispersion mask, which calculates positions for all wavelength bins along the dispersion axis \citep{Sandin10}. Furthermore, to remove instrumental scattered light from the CCD detector, we remove the spectra, interpolate the flux of the remaining background, and remove it from the raw spectra. 
 We absolute flux calibrate all the data using a spectral response function  calculated by comparing the observed stellar spectrum of the standard star and the spectrum from the \citep{Oke90} catalog of the corresponding star. The sky-subtracted stellar spectra is derived from the sum of fibers containing flux from the  standard star BD +33d2642.

  \subsubsection{Sky subtraction, relative flux calibration and data cubes}

The P3D package reduces the PPaK observations into calibrated row-stacked spectrum (RSS) images, from which we then subtract the sky contamination and flux-calibrate them.

 Due to the large spatial extent of M31 on the sky, none of the sky fibers, observed simultaneously with the field of view, could be used for sky subtraction. Dedicated sky fields were observed before and after every science field with 120\,s exposure times. The sky fields are processed following the same procedure as our science fields. We extract one median sky spectrum for each pointing of sky field observation and we linearly interpolate between sky exposures made before and after each science observations.  Such a simple interpolation is possible as the majority of sky emission features change slowly with time.  Because not all observing conditions were photometric and some observations ended during twilight, this interpolation technique failed for 23 of the pointings (out of a total of 50 pointings). In these cases, we assume that the median science spectrum across the field does not change significantly between dither positions.  We calculate a sky subtracted median of the science observation closest to the sky observation that appears least contaminated by clouds or twilight. Then we subtract this median science spectrum from the median observed spectrum in the remaining dither positions to recover a single sky spectrum.  Finally, we subtract this sky spectrum from each individual fiber spectrum in that dither position. 

To relative flux calibrate the RSS spectra we compare them with Sloan Digital Sky Survey (SDSS, \citealt{York00}) images of the same area (\citealt{Tempel11}). We use the positions and sizes of all fibers, apply the SDSS response functions to the spectra,  and compare resulting photometric fluxes with SDSS g and r band images. Then we scale each dither and combine them into pointings.

The final step was to combine the now flux-calibrated and sky subtracted RSS spectra for all pointings into a single 3D spectral data cube. To do this we combine each of the spectra onto a grid of 1\,arcsec$^2$ spaxels using a Delaunay linear triangulation \citep{Delaunay34} individually for each wavelength. Errors from the data and sky contribution (calculated by P3D) are propagated through the entire calibration process.

\begin{figure*}[t]
\centering
\includegraphics[width=0.9\linewidth]{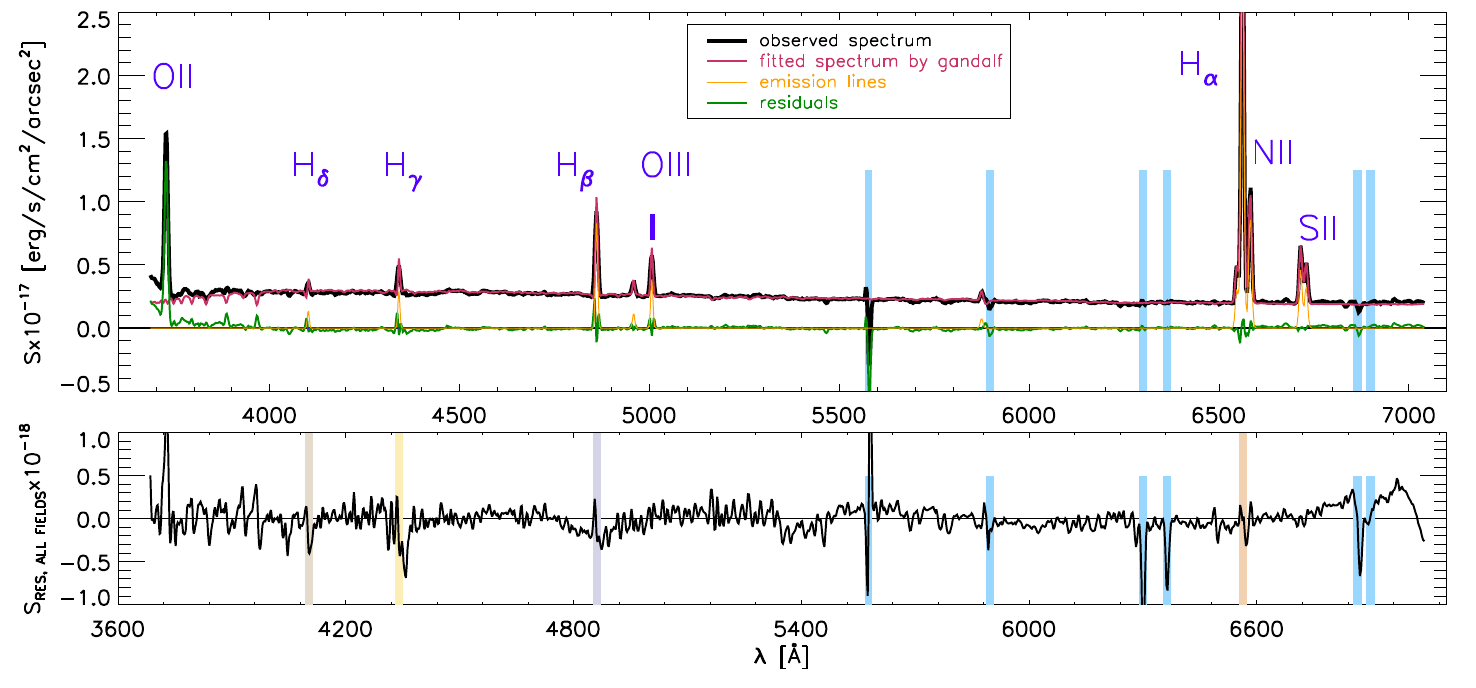}
\caption{ \textit{Upper panel}- Example  GANDALF spectrum fit for one spaxel in Field 1. The black line is the original scientific spectrum, the red line is the fitted spectrum (including  continuum and emission lines), the yellow lines show just the emission lines and the  green line shows the fit residuals. Bright emission lines are labelled, while bright omitted sky lines are shaded in blue. Note that the [OII]$\lambda$ 3727 emission line doublet is not fit with the other lines (see text for details). \textit{Lower panel}- Stacked residuals of all fields. Filled, coloured areas indicate the positions of Balmer lines, while the bright sky lines are shaded in blue.}
\label{fig:FigSpectrum}
\end{figure*}

\subsubsection{Line maps}

The optical galaxy spectrum contains  both stellar spectra (with continuum, emission and absorption lines) and emission lines from the ionized gas. We separate the nebular emission from the stellar spectra using the GANDALF\footnote{\textbf{G}as \textbf{And} \textbf{A}bsorption \textbf{L}ine \textbf{F}itting} software package\footnote{http://www-astro.physics.ox.ac.uk/~mxc/software/} version 1.5 \citep{Sarzi06}. GANDALF simultaneously fits both the emission lines and stellar continuum in an iterative approach. 

It fits the emission lines using the penalized pixel-fitting method (pPXF; \citealt{Cappellari04}). Each prominent emission line is fit by a Gaussian profile with the kinematics tied together (i.e.~$v$ and $\sigma$) but fluxes left free. To fit the stellar continuum we use template spectra taken from the \citet{Tremonti04} library of \citet{Bruzual03} simple stellar population (SSP) templates for a range of stellar ages (100 Myr to 12 Gyr) and metallicities ($ Z=0.004 $ and  $ 0.05  $). While SSP spectra may not represent the resolved stellar populations in our fields in M31, we chose to use the same templates as K13 for consistency. All templates are convolved to match the spectral resolution of our spectra. 

 We checked if using other SSP templates or stellar spectra (with higher spectral resolution and different stellar population) could change the results of our fitting and alter the underlying stellar absorption features. We used the MILES SSP and stellar spectra library templates (\citealt{Sanches06, Falcon11}) in fitting our spectra and found that the results do not show any  significant difference compared to using the  \citet{Tremonti04} library.

In fitting the continuum we also allow for a multiplicative third-order Legendre polynomial correction. This correction is necessary due to the poor flat-field correction in the blue part of the spectra caused by low CCD sensitivity, and to allow for stellar continuum attenuation by dust intrinsic to the fields of M31.

  Foreground extinction from the Milky Way is also accounted for in the spectral fitting. However this is considered to be uniform across the disk of M31 with an $A_{\rm V}=0.1705$\,mag in all fields (based on \citealt{Schlegel98} and \citealt{Schlafly11}).

The final data products from our fitting for each spaxel in our data cubes are: pure stellar continuum spectrum and fractional contribution of various SSP templates; multiplicative polynomial indicative of the intrinsic dust attenuation (and flat fielding corrections); gas velocity and gas velocity dispersion; and finally individual line amplitudes and fluxes for the most prominent emission lines. Identified  emission lines are: H$ \delta $, H$ \gamma $, H$ \beta $, [OIII]$ \lambda $4959\AA, [OIII]$ \lambda $5007\AA, [NII]$ \lambda $6548\AA, H$ \alpha $, [NII]$ \lambda $6583\AA, [SII]$ \lambda $6717\AA, [SII]$ \lambda $6731\AA.  

  Atmospheric optical emission lines (dominant around 5500   \AA{})  can cause problems with the sky subtraction. There is a weak Hg I 4358.34 \AA{} sky line (\citealt{Osterbrock92}, \citealt{Slanger00}) near the H$ \gamma $ line, which can also affect sky subtraction and fitting of the Balmer emission line. This effect is visible in the residuals near H$\gamma$ (Fig. \ref{fig:FigSpectrum}, seen as an absorption feature).  Additional contaminants like Earthshine and zodiacal light are removed from the spectra during sky subtraction as they exist as faint, extended features on the sky and  in the sky spectrum (\citealt{Reach97}). Geocoronal emission lines, spatially extended and slowly changing  with time, are removed by sky subtraction and do not have a significant effect on our Balmer line fluxes due to their narrowness (\citealt{Nossal01}, \citealt{Bishop04}, \citealt{Haffner03}). 

While the [OII]\,$\lambda3727$ \AA\,line doublet is detected in many of our spectra, the line is strongly affected by the poor sensitivity and subsequent calibration in the blue part of the spectrum, and hence it was not used in our analysis. Foreground stars (approximately 2-5 spaxels per field) were not removed from our data cubes, but spaxels affected by the stars are masked during our GANDALF spectral analysis.

The median $ 3\sigma $ sensitivities of H$ \alpha $ and H$ \beta $   in all fields are $ 7.6 \times 10^{-18} $ and $ 4.9 \times 10^{-18}$ erg\,s$ ^{-1} $cm$ ^{-2} $arcsec$ ^{-2} $,  respectively. However, averaged or median $ 3\sigma $ sensitivities of Balmer lines do not determine whether the data from spaxels will be shown or be excluded from the following diagrams and measurements. Data are included only if a line's amplitude is above the $ 3\sigma $ noise on the continuum. GANDALF calculates the noise on the continuum as the standard deviation of the residuals for the entire wavelength range, while uncertainties in the line fits as the amplitude over noise values (AoNs; \citealt{Sarzi06}).  Fig. \ref{fig:FigSpectrum} presents an example of the spectrum fitting for one spaxel in  Field 1.

\subsection{Far-IR data and dust column densities}

Far-IR (FIR) emission is a good tracer of the amount of the dust in the line of sight (\citealt{Williams82, Neugebauer84}). Therefore we use the far-IR data observed by the PACS and SPIRE camera on the ESA \textit{Herschel Space Observatory} (\citealt{Griffin10}, \citealt{Pilbratt10}). The dust mass surface density map of M31 presented by \citet{Draine14}, used for the comparison with attenuation maps, was  determined by  fitting the spectral energy distribution (SED) of the near and far IR emission (\citealt{M31Groves12}) with the \citet{DraineLi07} dust  model. The \citet{DraineLi07} model specifies the dust characteristics like distribution of grain sizes, frequency-dependent opacity, fraction of dust in polycyclic aromatic hydrocarbons (PAHs) and dust column densities. The models were determined by calculating emission spectra (in near-IR, FIR and sub-millimeter) and reproducing extinction curves for different abundances of small PAHs and various dust mixtures heated by different starlight intensities. 
 
 The resulting dust mass surface density map ($ \Sigma_{\rm dust} $) has an effective 24.9$ '' $ Gaussian PSF (matched to SPIRE 350\,$ \mu m $ resolution).  In order to compare our final attenuation maps with the dust mass surface density maps, we convolved our cubes from the effective PPaK resolution of 2.7$ '' $  into the SPIRE 24.9$ '' $ resolution. To convolve the data we use the kernels and the routine described in \citet{Aniano11}. 
 
 We convolve the data cubes by splitting them into images for each wavelength bin, and convolving each image separately. Before the convolution process, we add extrapolated values to the area outside the edges of the image, convolve the image, and then replaces those areas with blanks after the process. Then we reassemble the cubes to the SPIRE 350\,$ \mu m $ image grid in order to compare the maps. After the convolution, we perform the same fitting routine and spectral analysis on the convolved images as described in Section 2.1.3. 
 
\begin{figure*}[ht!]
\begin{center}
\includegraphics[width=0.9\linewidth]{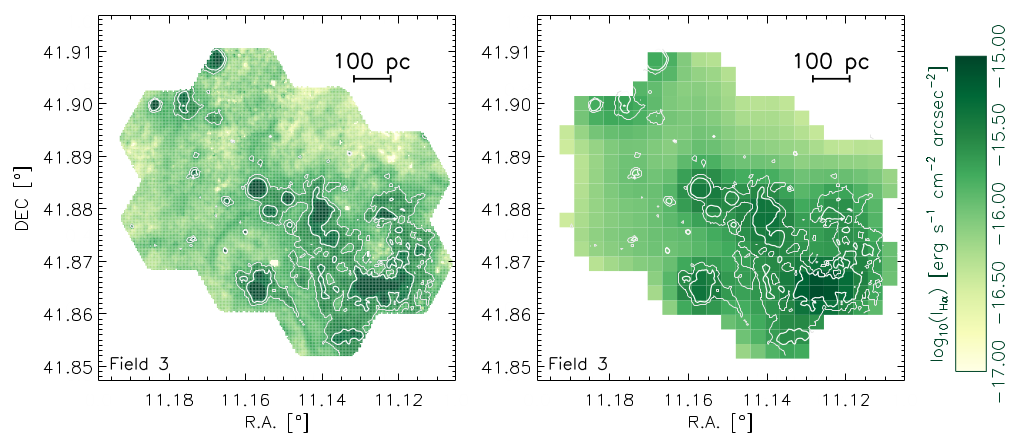}
\caption{ An example of convolving the   H$ \alpha $  intensity images  using Field 3. On the left is the un-convolved image at the native PPaK resolution of 2.7$ '' $. Contours represent the H$ \alpha $ intensities of $ 1.5\times 10^{-16} $ (thin) and  $ 3\times 10^{-16}  $(thick) erg\,s$ ^{-1} $cm$ ^{-2} $arcsec$ ^{-2} $. Flux from filamentary diffuse ionized gas can be clearly seen, spanning the area between HII regions. On the  right is the convolved image at the SPIRE 350\,$ \mu m$, 25$ '' $ resolution. Due to the increased S/N in the convolved images, some areas that are not robustly detected in the native images are included in the convolved ones (see the white contours in the Field). These features are not significant in the native map, yet clear emission lines are seen in every spaxel in the convolved image. }
\label{fig:Fig03}
\end{center}
\end{figure*}

 This convolution technique changes the Balmer lines intensities (up to 30\%) on the edges of the data cubes, depending on the position of bright HII regions. It affects both Balmer lines simultaneously, which results in only small changes in A$_{\rm V}$. If we compare A$_{\rm V}$ maps derived from the convolved data cubes and smoothed (but not convolved) cube, changes in A$_{\rm V}$ can be up to 0.3 mag for the bright regions. The effects of foreground stars  are minimized due to the convolution process. Fig. \ref{fig:Fig03} shows an example of our un-convolved and  convolved maps of H$ \alpha $  emission within Field 3.

\section{Results}

   With the resulting data cubes and analyzed spectra, we measure the optical attenuation (A$_{\rm V}$) and compare it with the dust mass surface density ($ \Sigma_{\rm dust} $). In the following subsections we describe the calculation of A$_{\rm V}$, show the resulting maps and compare the A$_{\rm V}$ and $ \Sigma_{\rm dust} $ maps.

\subsection{Attenuation maps}

   Due to reddening by dust (correlated with the extinction) the  Balmer  line ratios (known as Balmer decrements) are altered from their intrinsic ratios.  The total V-band extinction (A$ _{\rm V} $) is related to nebular reddening E$_{\rm B-V}$ by:    
  \begin{equation}
 A_{\rm V}\equiv R_{\rm V}E_{\rm B-V} ,
 \label{eq:Eq01}
 \end{equation}
where $R_{\rm V}$  is the selective extinction. R$ _{\rm V} $ depends on the physical characteristics of the extinguishing dust grains. In the diffuse ISM of the Milky Way, $R_{\rm V}$ has an average value of 3.1, which is the value typically assumed for massive star-forming galaxies (\citealt{Schultz75}, \citealt{Cardelli89}, \citealt{Calzetti00}). We assume the same R$_{\rm V}$ value for M31 as our fields are at similar galactic radii as the Sun (at 0.3-0.6 R$_{\rm 25}$; \citealt{Bigiel11}) and with similar metallicities (\citealt{Zurita12}, \citealt{Draine14}). The reddening between two lines (F$ _{1} $ and F$ _{2} $)  is calculated as \citep{Calzetti94}:
 \begin{equation}
 E_{\rm B-V}=\frac{2.5}{k_{2}-k_{1}}\log_{10}\left( \frac{F_{1}/F_{2}}{R_{int}}\right) ,
 \label{eq:Eq02} 
  \end{equation}
 where $ k $ is the extinction as a function of wavelength for the corresponding lines.  Here we use the extinction curve from \citet{Cardelli89}. By using the \citet{Calzetti00} attenuation curve instead, the resulting inferred A$_{\rm V}$  would decrease by 9\%. That systematic shift should be kept in mind when comparing our results with those found by K13, where they used the \citet{Calzetti00} curve. The difference between those two curves is because the  \citet{Cardelli89} curve solely accounts for foreground extinction while the \citet{Calzetti00} curve takes into account geometrical effects on attenuation.
 
 We assume an intrinsic flux ratio of  R$_{int}=\mathrm{H\alpha}/\mathrm{H\beta}=2.86$, corresponding to an ionized gas temperature of T$ \approx 10^{4} $ (assuming case B recombination; \citealt{Miller74}, \citealt{Osterbrock74}, \citealt{Osterbrock92},\citealt{Osterbrock06}).

\begin{figure*}[ht!]
\centering
\includegraphics[width=0.9\linewidth]{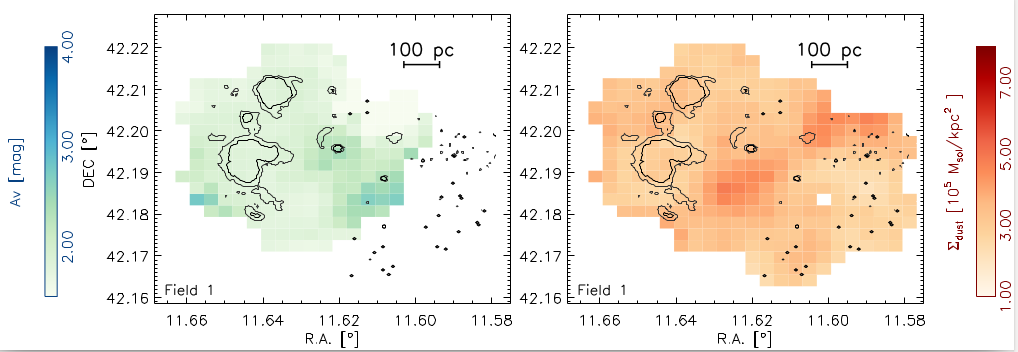}
\centering
\includegraphics[width=0.9\linewidth]{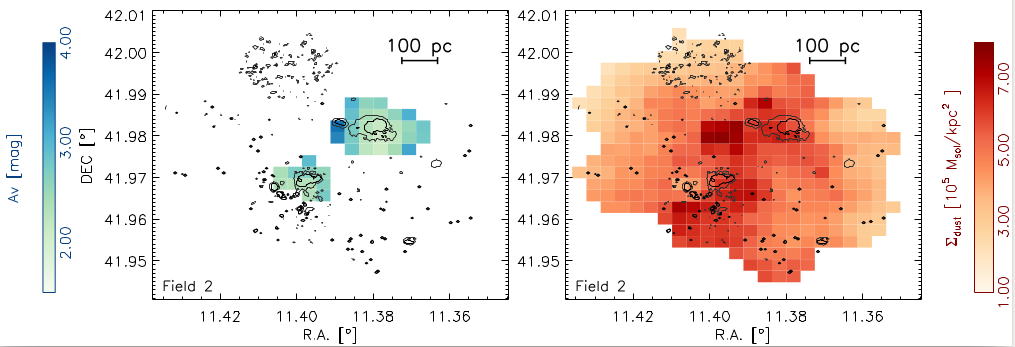}
\caption{The attenuation (A$ _{\rm V} $, left) and dust mass surface density maps ($ \Sigma_{\rm dust} $, right) for Fields 1 (top) and  2 (bottom). Black contours represent the H$ \alpha $ intensities of $ 1.5\times 10^{-16} $ (thin) and  $ 3\times 10^{-16}  $(thick) erg\,s$ ^{-1} $cm$ ^{-2} $arcsec$ ^{-2} $. Only spaxels with AoN of Balmer lines greater than 3 are shown. All maps are at the same scale, allowing for direct comparison between the fields. Field 1 has more uniform attenuation throughout the field. Similarly, that is matched by a uniform dust distribution that has a lower surface density compared to the other fields. One noticeable feature for Field 1 as compared to the other fields is the slight anti-correlation between the positions of the HII regions and dust. }
\label{fig:Fig04}
\end{figure*}

\begin{figure*}[ht!]
\centering
\includegraphics[width=0.9\linewidth]{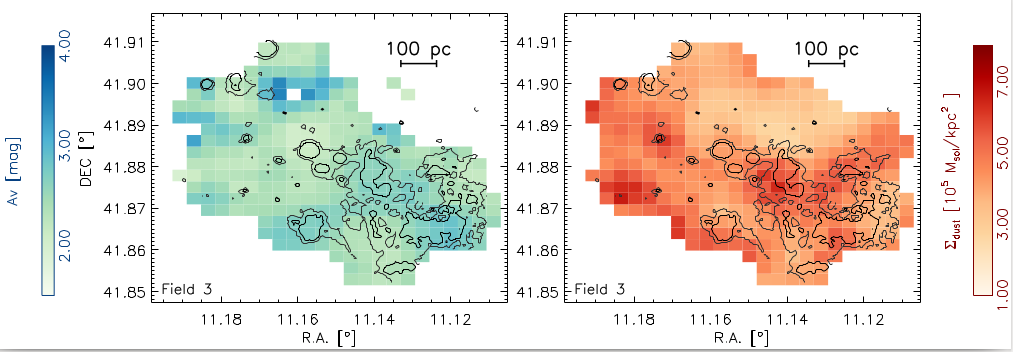}
\centering
\includegraphics[width=0.9\linewidth]{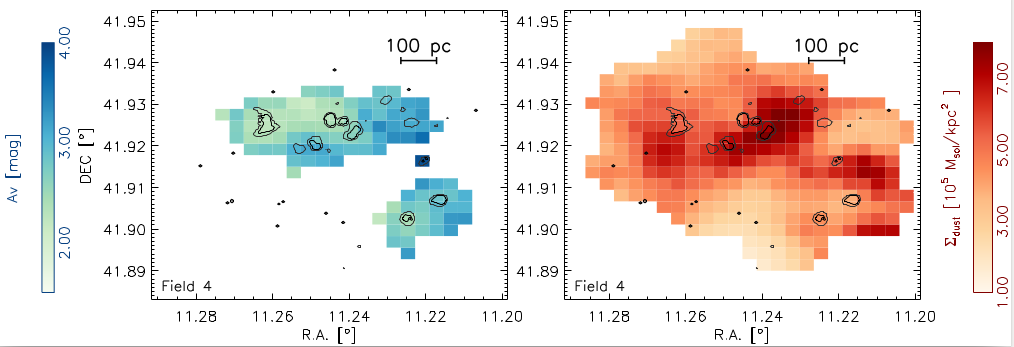}
\centering
\includegraphics[width=0.9\linewidth]{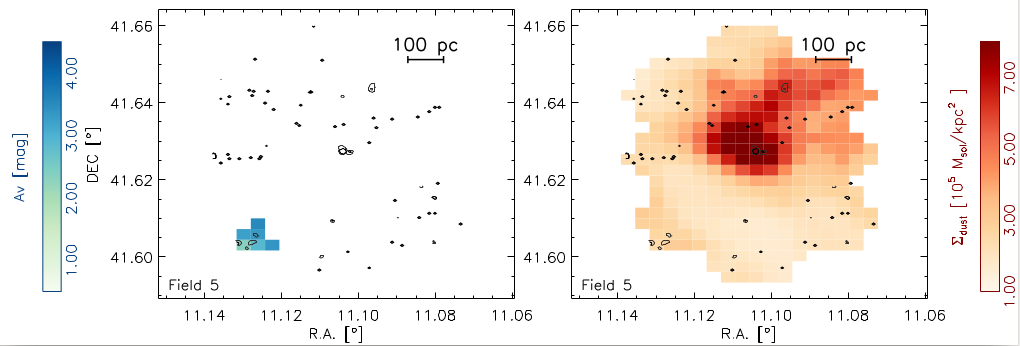}
\caption{ The attenuation (A$ _{\rm V} $, left) and dust mass surface density maps ($ \Sigma_{\rm dust} $, right) for Fields 3 (top), 4 (middle) and  5 (bottom). Contours and scale are as in Figure 4. The attenuation spans a larger range of values compared to Field 1. In general, HII regions are situated mostly in or near the regions of higher dust mass surface density. Due to the low flux and hence S/N of the emission lines in Field 5, there are only four pixels in the convolved map with robust detections. }
\label{fig:Fig05}
\end{figure*}

   We show attenuation (A$ _{\rm V} $) maps and the dust mass surface density maps ($ \Sigma_{\rm dust} $) in Fig. \ref{fig:Fig04} and Fig. \ref{fig:Fig05}. For all our fields contours show the H$ \alpha $ intensities, tracing the position of the HII regions. All maps are at the same scales, which offers a direct comparison between the fields. A$ _{\rm V} $ spans values between 1.5-4.5 mag in our fields. Similar results are observed by \citet{Sanders12}, where observed HII regions in M31  show values between 1 and 5 mag. The fact that there are no M31 data points with attenuation lower than A$ _{\rm V}=1 $ is an effect of targeting dense spiral arm regions in M31, which biases us to regions of high $ \Sigma_{\rm dust} $. Were our fields larger and  included less dusty regions we would expect our maps to have more data points with A$ _{\rm V}<1 $.

\clearpage

\begin{figure*}[ht!]
\centering
\includegraphics[width=0.7\linewidth]{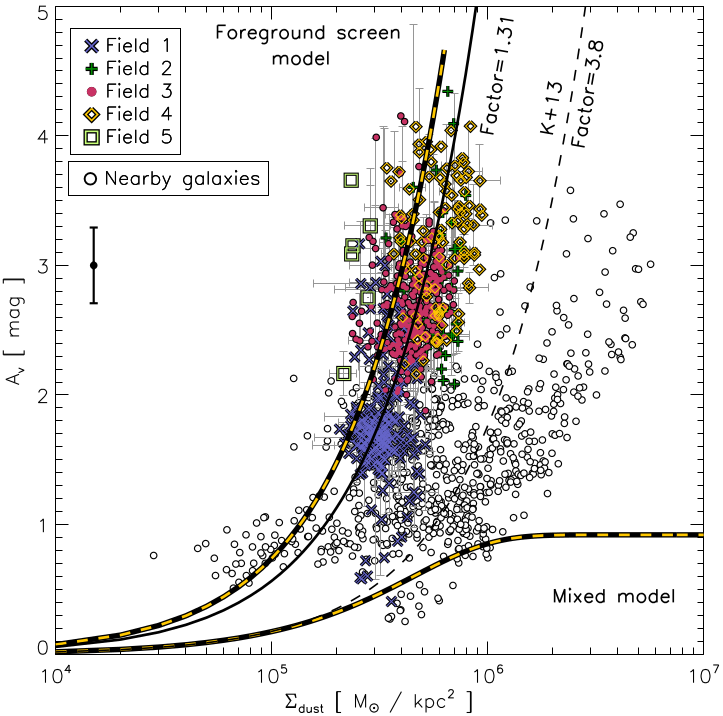}
\caption{ Attenuation (derived from the  H$ \alpha $/H$ \beta $ ratio) compared to the dust mass surface density. Foreground screen and mixed models \citep{Calzetti94} are represented by thick, black/yellow lines. Various symbols represent data points from our five fields in M31.  Data from K13 for nearby galaxies are shown in the background as empty circles.  A foreground screen model decreased by a  factor of 3.8 (from K13) and 1.35 (best fit for M31) are also shown. The median value of the M31  uncertainties of attenuation values are shown on the left side of the diagram.}
\label{fig:FigMain}
\end{figure*}

    In general, we find that the HII regions are situated near or in the regions of higher dust mass surface densities. This is expected if the dust traces the regions of high gas density where new stars are formed.  Contrary to other fields, in Field 1 the HII regions are situated in less attenuated and less dusty areas. We explain this as an effect of stellar feedback from more evolved HII regions that have already destroyed dusty birth cloud.

  We exclude from the following analysis and maps regions where the Balmer lines do not exceed a threshold of AoN $ \gtrsim $ 3 (corresponding to S/N $ \gtrsim $ 3).  Due to the low flux and hence S/N of the emission in Field 5, there are few spaxels that exceed this threshold. This causes some statistical difficulties in the analysis of this field. There is also a possibility that a young HII region is buried in the dense cloud, seen in the center of Field 5 (Fig. \ref{fig:Fig05}).

  Our calculated attenuation  depends on  the physical condition of the ionized gas (which can cause different intrinsic line ratios) and on the dust composition (resulting in different values of $R_{\rm V}$), both of which affect the extinction curve. Therefore, the attenuation values of some data points can be different due to intrinsically different physical condition in those regions. However, these effects are presumed to be small relative to the real variation in A$_{\rm V}$ due to the dust distribution.

\subsection{ A$ _{\rm V} $ vs. $ \Sigma_{\rm dust} $}

 The main goal of this paper is to compare the dust in M31 as determined via two independent methods; the attenuation derived from the Balmer decrement, and the dust mass surface density derived from IR photometry. This follows on from the work of K13 who found a relation between these measures within eight nearby galaxies, but with large scatter between galaxies. 

The dust attenuation and dust mass surface density are connected via the distribution of dust. We consider here two simplistic models of the dust distribution, derived from \citet{Caplan86} and \citet{Calzetti94}. The Calzetti models predict the correlation between attenuation and the dust mass surface density for five different spatial distributions of the dust. The two major models that are used for this work are the `foreground screen' model and the `mixed' model  (in \citealt{Calzetti94} these are referred to as the ``Uniform dust screen'' and the ``Internal dust'' models). The remaining models are variation on those two extremes.

    The `foreground screen' model describes a situation where all the dust sits in a smooth, uniform screen between the emitter  and the observer (the dust is not mixed with the sources of radiation). Assuming the dust model from \citet{DraineLi07} and the dust-to-gas ratio (DGR) from \citet{Draine14}, the attenuation is linearly related to  $\Sigma_{\rm dust}$  via:

 \begin{equation}
 A_{\rm V}^{\rm screen}=0.74\cdot\frac{\Sigma_{\rm dust}} {10^{5} \rm M_{\odot}\rm kpc^{-2}}  [mag].
 \label{eq:Eq03}
 \end{equation}       
Given our assumptions, this equation provides a  theoretical upper limit on the attenuation possible at a given dust mass surface density.

 The  `mixed' model assumes the dust is uniformly distributed with the sources of radiation. Therefore the attenuation in the mixed model is much lower than in the foreground screen model.
The resulting mixed model attenuation, assuming isotropic scattering and the same connection with dust mass as in equation \ref{eq:Eq03}, is (based on \citealt{Calzetti94}):

\begin{equation}
A_{\rm V}^{\rm mixed}=-2.5\log_{10}(\gamma_{\rm V})[{\rm mag}], 
\label{eq:Eq04}
\end{equation}

where
\begin{equation}
\gamma_{\rm V}=\frac{ 1-e^{0.57 A_{\rm V}^{\rm screen}} }{ 0.57 A_{\rm V}^{\rm screen}}.
\label{eq:Eq05}
\end{equation}

 $ \gamma $ functionally limits the value of the optical depth in the optically thick medium \citep{Calzetti94}. The screen model correlates linearly with the dust column,  while the mixed model saturates and for high dust mass surface densities yields a moderate expected attenuation (Fig. \ref{fig:FigMain}).

Fig. \ref{fig:FigMain} shows a comparison of the dust mass surface densities and the attenuation for M31 and for the nearby galaxies observed by K13, together with the two Calzetti models. While the data from K13 span the area between the foreground screen and mixed models, the more resolved M31 data follows more closely a foreground screen model. No clear trends are seen within each Field of M31, but when the fields are considered together the data fits very well the foreground screen model. We find the correlation between A$ _{\rm V} $ and $ \Sigma_{\rm dust} $ well fit  by scaling down the foreground screen model by  a factor of 3.8 for the nearby galaxies (K13) and  1.35 for M31. 
 
We see some regions in  M31 that have larger attenuation than is expected from our foreground screen model (Eq. \ref{eq:Eq03}). Approximately 25\% of the data points are above the foreground screen model, or 3\% of the data  if we take into account their corresponding 1$ \sigma $ errors.  These high A$_{V}$ regions can be explained by: (1)  'clumpiness' in the dust below our resolution that is affecting our measured $ \Sigma_{\rm dust} $,  (2) an underestimation of the dust mass surface density,  (3) poor calibration of the optical spectra,  (4) values of  A$_{\rm V}/\Sigma_{\rm dust} $ different from what we have assumed, (5)  variation in the extinction curve k$(\lambda)$ that affect  R$ _{\rm V} $ and thus A$_{\rm V}$. 

 If an area with low $ \Sigma_{\rm dust} $ has locally a high density clump of dust covering the HII region, averaging the dust surface density due to the low spatial resolution of our observations could misleadingly show a low $ \Sigma_{\rm dust} $ and a high A$ _{\rm V} $. However, when comparing the value of A$ _{\rm V} $ around HII regions at PPaK 2.7$ '' $ resolution  ($ \sim 10$ pc) , we find that A$ _{\rm V} $ does not change rapidly with different aperture sizes.

   Previous work has tested the accuracy of estimating dust masses using the \citet{DraineLi07} model (\citealt{Alton04, Dasyra05, Galametz12, Aniano12Dust, Dalcanton15, Planck16b}).  Most recently,\citet{Dalcanton15} and \citet{Planck16b} measured dust column density within the Milky Way and M31 by measuring the extinction of the light from background sources.  Both studies concluded that the \citet{DraineLi07} dust model may overestimate the mass of the dust by a factor of $ \sim $2.5. Comparison with independent far-IR observations have shown that this offset is not due to uncertainties in the Herschel photometry \citep{Verstappen13, Planck16b}. 

\citet{Planck16b} suggest that this offset is dependent upon the heating radiation field intensity. In the Appendix we explore the effects of renormalizing $ \Sigma_{\rm dust} $ in M31 and the K13 (Appendix A; Fig. \ref{fig:AppendixA}). Based on this we would expect that such a normalization would only further the disagreement between the K13 galaxies and M31. However, as this renormalization is not well calibrated in the high interstellar radiation field regime (corresponding to most of the K13 regions),  we do not include the renormalization in the following analysis and figures.
  
 An additional effect that can  play a role in our derived attenuation is the 'mid-plane' effect, as we expect half of the dust to be situated behind the ionizing sources. That can partly explain the offset of M31 data from the  `foreground screen' model line.  This effect should be strongest in the case where the scale height of the ionized gas is much smaller than the scale height of the dust. However, in the case of similar scale heights, this effect should play a much smaller role.       
  
  Given our confidence in our spectrophotometric calibration, we explain higher attenuation values of some regions as a result of possible variations in the extinction law tied to variations in the dust properties, which we expect are related to variations in our assumed DGR and R$ _{\rm V} $.

\section{Discussion}

Our main result (Fig. \ref{fig:FigMain}) shows that M31 more closely follows a foreground screen model than a mixed model.  This is different from what K13 found in nearby galaxies. In this section, we consider various factors that potentially explain the differences between these results. We examine the effect of physical resolution on attenuation (Sec. \ref{Sec: 4.1 Scales}). We test the effects of  a spatially extended ionized gas component  on attenuation (Sec. \ref{Sec: 4.2 Additional flux}). We associate this additional gas with diffuse ionized gas (DIG) and discuss the effects of different scale heights of dust, HII regions and DIG on A$ _{\rm V} $ (Sec. \ref{Sec: 4.2 Additional flux}, \ref{Sec: 4.3 SIIHa vs Av} and \ref{Sec: 4.3 Toys}). Finally, we explain the varying vertical distribution of dust and gas in M31 and the K13 galaxies by observations at different galactic radii (Sec. \ref{Sec: 4.4 Env}).

\subsection{The impact of physical resolution}
\label{Sec: 4.1 Scales}

While K13 probed spatial scales between $\sim 0.3$\,kpc to $\sim 2$\,kpc, the proximity of  M31 means that the SPIRE 350\,$ \mu m$ physical resolution is $\sim100$\,pc. The  fields in M31 are located within the most dense and dusty spiral arms and cover only a small fraction of the galaxy. Therefore it is possible that our M31 results are  biased by dusty, star-forming regions where a foreground screen model more closely represents the dust distribution.  The larger physical sizes of the regions sampled in K13 can probe both regions where a foreground screen model is more representative and regions where a mixed model more closely matches reality. This yields a result where together all regions show a mixture between the two models. K13 searched for a correlation between  A$ _{\rm V} $ - $ \Sigma_{\rm dust} $ slope and different spatial scales, but the changes in  slope were not significant.

   To directly compare our result with K13 (whose best spatial resolution is $ \approx $300\,pc), we smoothed our data (Fig. \ref{fig:FigScales}) to 65$ '' $ resolution (corresponding to spatial scales of $ \approx $260\,pc) and find no difference in the slope of the correlation. Moreover, we also integrate all spectra from each field into a single data point and re-extract the line fluxes to determine the average attenuation for each field. Resulting integrated field data points span spatial scales of around 0.6\,kpc$ \times $0.9\,kpc. We then compare this value to the average dust mass surface density for the field. Integrated field data points are shown as circles in Fig. \ref{fig:FigScales}, and it is clear that even at these scales we see the same relation between attenuation and dust mass surface density. Some fields are not presented, due to the low number and brightness of HII regions. As H$ \beta $ line is faint, this results in low AoN values for those fields. 
   
The regions in K13 target wide areas (covering both spiral arms and inter-arm regions) and can consist of regions with or without low-brightness HII regions. It is possible that the dust probed in some of those regions is heated by an old stellar population and not solely by HII regions. However, as the \citet{DraineLi07} model explicitly takes into account starlight heating, the modeled dust mass should not suffer from any bias in either K13 or our results on any of the scales considered. Using the close alignment of disk galaxies in front of early-types, \citet{Holwerda13} found that inter-arm regions contain less dust relative to spiral arms. While K13 included regions with a broad range of dust mass surface densities, the regions with weak H$ \alpha $ flux (where dust may be predominantly heated by the old stellar population) are generally removed by the S/N criteria in their work. Therefore we conclude that the difference between our results and their is not related to the treatment of the inter-arm regions. However, the integrated fields of M31, which have a similar spatial resolution to K13 ones, have star-formation rate surface densities ($ \Sigma_{\rm SFR} $)  that are a factor of 10 lower. This makes a direct comparison between fields at fixed $ \Sigma_{\rm SFR} $  impossible. The mean  $ \Sigma_{\rm SFR} $ in M31 is around $  \sim 0.01$ M$ _{\odot} $\,yr$ ^{-1} $\,kpc$ ^{-2} $, while for the K13 galaxies $\Sigma_{\rm SFR}$ spans a $  0.03-10 $ M$ _{\odot} $\,yr$ ^{-1} $\,kpc$ ^{-2} $  range (Appendix B; Fig. \ref{fig:AppendixB}).

\subsection{Effects of an additional component to the dust/gas distribution model}
\label{Sec: 4.2 Additional flux}

Another possibility to explain the difference in the A$ _{\rm V} $ - $ \Sigma_{\rm dust} $ relation between M31 and the K13 galaxies is  that, in the K13 galaxies, some fraction of the ionized gas resides outside the dust disk. As this emission would not be extinguished, therefore lowering the observed A$ _{\rm V} $. In this section, we propose the diffuse ionized gas (DIG) as a candidate for this gas outside the dust disk. We also explain why the previously used \citet{Calzetti94} model of dust/gas distribution (combining only HII regions, stars and dust) need an additional diffuse and spatially extended component.

\subsubsection{ Additional flux from non-attenuated gas}
\label{Sec: 4.2.1}

The \citet{Calzetti94} model for the dust/gas distribution combine only dust and stars (that ionize the gas within the HII regions). That model neglects the possibility of additional  flux from ionized gas spatially even more extended than the dust. In the scenario where additional gas presides outside the dust disk, the observed attenuation would be lower than in the case of the \citet{Calzetti94} model, even for the same observed amount of dust. If we assume that the M31 regions have all of the ionized gas  embedded in the dust disk, we explore how flux from a non-attenuated ionized gas, which may  lie outside the dust disk, would effect the A$ _{\rm V} $ - $ \Sigma_{\rm dust} $ relation.      
 
 In Fig. \ref{fig:FigDiff} we investigate how large the contribution from a non-attenuated ionized gas component above the dust disk would need to be in order for the M31 data to follow the trend of K13. H$ \alpha $ flux from that non-attenuated ionized gas, F$^{\rm non-ext} $, is presented as a  percentage of the total observed flux, F$^{\rm tot} $. 
 For M31, we  assume that no ionized gas is outside the dust disk. The left panel in Fig.  \ref{fig:FigDiff} shows the impact on our integrated fields in M31 when we add flux from such an additional ionized gas outside the dust disk.  Similarly, the right panel shows the impact of that additional ionized gas on the foreground screen model. Even a small amount of flux from the added ionized gas lowers the attenuation enough that we observe values similar to those found by K13.

 If  30\%-60\% of  the observed total H$ \alpha $ flux arises from  gas which sits outside the dust disk, we can recover the relation observed by K13 within nearby galaxies. This provides a better qualitative fit  than simply scaling the foreground screen model by a  factor of 3.8, as suggested by K13 (Fig. \ref{fig:FigDiff}, right panel). When the contribution from the additional ionized gas component to the emission is large, our model drops even below the `mixed' model curve. This is because, unlike the Calzetti model, we are including a component with no attenuation at all in the model. At extreme contributions we would only see the non-attenuated component and measure little to no total attenuation. 
 
 It is not expected that the ionzied gas directly associated with HII regions will span large spatial scales, as typical HII region sizes are less than 100 pc \citep{Azimlu11}. However, a good candidate for the extended ionized gas component we propose here is the diffuse ionized gas.

 \subsubsection{ Diffuse ionized gas (DIG) as an additional component }
 \label{Sec: 4.2.2}

Diffuse ionized gas (DIG, also known as the Warm Ionized Medium or WIM; \citealt{Reynolds71, Walterbos94, Greenawalt98, Wang99, Oey07, Haffner09}) extends within and outside the galactic disk. Unlike HII regions and most of the dust, which reside within a thin disk and span scale heights of only $ \approx $100\,pc, the DIG can extend above and below the disk out to kpc scales, similar to the thick disk component (\citealt{Reynolds84, Haffner09, Bocchio16}).  Previous studies have shown that the scale height, location and brightness of the DIG follows the brightness and location of star forming regions (\citealt{Dettmar90}, \citealt{Dettmar92}, \citealt{Ferguson96}, \citealt{Ferguson98}, \citealt{Rand98}, \citealt{Wang99}, \citealt{Collins00}, \citealt{Rand05}, \citealt{Heald06}, \citealt{Oey07}).   The DIG can be a good candidate for the extended ionized gas, which may lie outside the dust disk and may affect the attenuation.  Due to the $\sim$kpc physical resolution observed by K13, they were unable to distinguish the emission arising from HII regions and the DIG.

Physically, the DIG is warmer and less dense than gas in HII regions. In the DIG regions, the [SII]$(\lambda6717+\lambda6731)/$H$\alpha $ ratio is higher ($>$ 0.4, while for the HII regions it is usually $<$ 0.2; \citealt{Minter98, Haffner09}). The origin of ionization in the DIG is speculative, but is likely due to a combination of supernova shocks, turbulent dissipation, leaked radiation from nearby OB stars, additional photons provided along channels in the neutral gas, and heating by cosmic rays or dust grains (\citealt{Reynolds90, ReynoldsCox92, Minter98, Madsen06, Haffner09, Barnes14, Barnes15, Ascasibar16}). 

 In M31, \citet{Walterbos94} found that after masking all HII regions the DIG flux contributes on average around 40 \%  of the total H$ \alpha $  flux (or 20 \% after extinction corrections). This percentage varies between the observed fields in M31 (\citealt{Walterbos94}) and  is observed to vary greatly between galaxies \citep{Haffner09}.  

  The DIG fraction we detect in our fields in M31 agrees with the results of \citet{Walterbos94}; after masking HII regions using an intensity threshold of $ 2\times10^{-16} $erg\,s$ ^{-1} $cm$ ^{-2} $arcsec$ ^{-2} $ (as used in \citealt{Walterbos94}),  the contribution of DIG flux to the total flux of the fields varies between 40\% (Fields 1 and 3) and 70\% (Fields 2 and 4). Most of the emission in Field 5 is  composed of DIG flux (based on this threshold value). The line ratio of [SII]/H$\alpha $ is $ \approx $0.5  throughout   our fields and the fields observed in \citet{Walterbos94} (which are almost 5 times larger). 
  
   The intrinsic  H$\alpha$/H$\beta$ ratio of the DIG is likely to be higher given the higher temperature, but we assume it to be 2.86 for simplicity.  Using ITERA\footnote{\textbf{I}DL \textbf{T}ool for \textbf{E}mission-line \textbf{R}atio \textbf{A}nalysis} \citep{Groves10}, we conclude that the intrinsic Balmer lines ratio of DIG can have values between 2.75 and 3.1, depending on the exact temperature, ionization parameter, magnetic field strength and density. Even if the intrinsic ratio is an extreme value of 3.1, the attenuation values would be lower by only approximately 0.3-0.5 mag from those currently shown in Fig. \ref{fig:FigMain}, which would not dramatically change our results.  Therefore, in the following we assume the same intrinsic ratio while calculating the A$ _{\rm V} $ of the DIG. 

\subsubsection{Effects of different dust-HII-DIG distributions on A$ _{\rm V} $ }
\label{Sec: 4.2.3}

 The relative  distribution and differences in scale heights between HII, DIG, and dust can change our derived attenuation even if the relative amounts stay the same.  For example, if both components of ionized gas (DIG and HII regions) have smaller scale heights compared to the dust, the derived A$ _{\rm V} $ of both components would only correlate with $ \Sigma_{\rm dust} $. If the DIG presides outside the dust (i.e. has a larger scale height), the average A$_V$ of the ionized gas (combining both HII region and DIG emission) will be lower for the same $\Sigma_{\rm dust}$. That scenario is equivalent to the one tested in Fig. \ref{fig:FigDiff} (Sec. \ref{Sec: 4.2.1}).    
 
   Multiple results indicate the need for more complex dust distribution models within galaxies in order to match simulations and observations (\citealt{Wong02}, \citealt{Popescu11}, \citealt{Viaene17}). \citet{Wong02} proposed a ``hybrid'' model where dust behaves differently in region located in the center/inner and outer disk of galaxies.  In the central regions of galaxies, where molecular gas dominates, the dust is more dense and has a smaller scale-height. In the outer  regions of galaxies, where the HI gas dominates, the dust is more diffuse and vertically  more extended.  
 This complexity in the dust scale height  may explain the difference between M31 and the galaxies studied by K13. 
 
  If we assume that in M31 the ionized gas (HII and DIG regions) presides within the dust layer (partly mixed with the dust and partly obscured by a dust screen), then the attenuation would follow the slope of the foreground screen model. For the nearby galaxies in K13, we propose that the DIG presides outside the dust and that dust and gas are not mixed. In that case, the non-attenuated flux from the DIG contributes to the already attenuated flux from HII regions (which lie within the dust), and lowers the overall attenuation. It is unlikely that there is no attenuation at all of the DIG component as it is probably mixed with the diffuse dust component in the thick disk (\citealt{Bocchio16}, \citealt{Howk12}). However, for simplicity we assume that most of the DIG in nearby galaxies do not mix is external to the dust.

\begin{figure*}[ht!]
\centering
\includegraphics[width=0.5\linewidth]{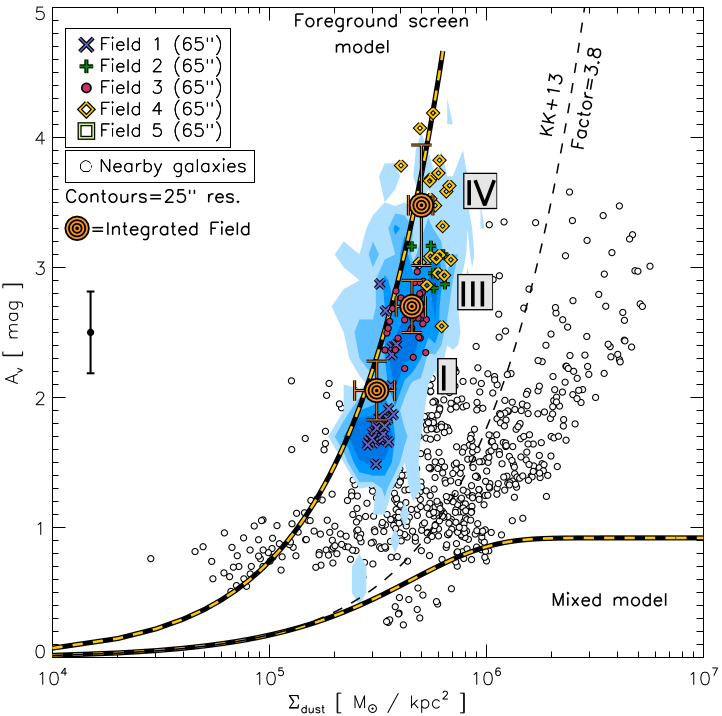}
\caption{ A$ _{\rm V} $ vs. $ \Sigma_{\rm dust} $  for our M31 fields convolved to different spatial scales. Contours show the data from Fig. \ref{fig:FigMain} at 24.9$ '' $ scales( $ \sim $100\,pc), the various symbols represent the data at  65$ '' $ scales ($ \approx $260\,pc), while the big circles indicate results from integrating each Field (as labelled) spanning  spatial scales of 0.6\,kpc$ \times $0.9\,kpc.  Due to the low AoN of H$ \beta $, only three fields can be  shown.  No obvious change  in the $A_{\rm V}$ versus $\Sigma_{\rm dust}$ relation is apparent between the different resolutions.  1$ \sigma $ uncertainty error bars for the integrated fields and the median uncertainty of the convolved data (shown in the left corner) are presented. Systematic uncertainties in the spectral fitting dominate over the random instrumental uncertainties.  }
\label{fig:FigScales}
\end{figure*}

\begin{figure*}[ht!]
\centering
\includegraphics[width=0.95\linewidth]{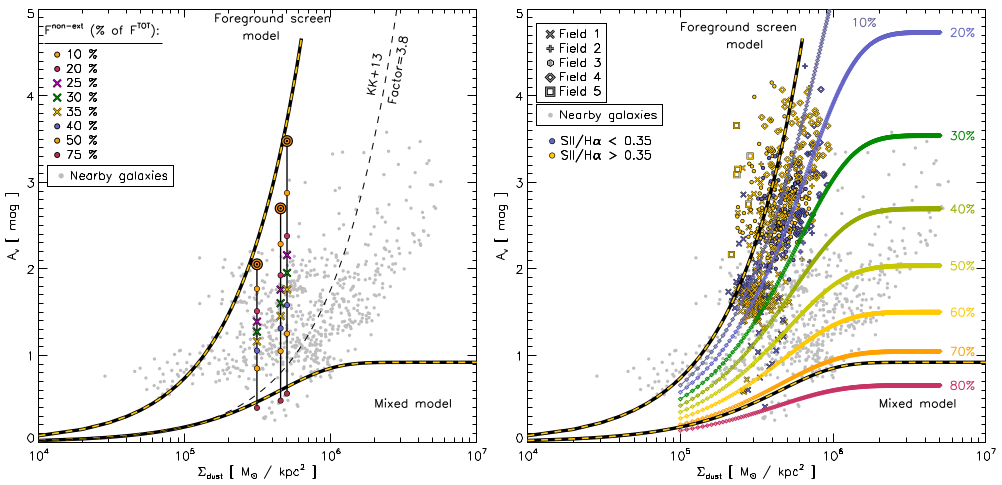}
\caption{The impact on attenuation when flux from non-attenuated  ionized gas (F$^{\rm non-ext} $)  is added outside the dust disk. F$^{\rm non-ext} $ is shown  as a  percentage of the total flux  F$^{\rm tot} $. The left panel shows the impact on A$ _{\rm V} $ for integrated fields in M31 when we add non-attenuated flux. The right panel shows the impact of unattenuated emission on the foreground screen model. The lines on the right panel better reproduce the trend and slope of the K13 data than the  scaling factor of 3.8 for the foreground screen model proposed by K13. Some lines are situated below the mixed model limit predicted by \citet{Calzetti94} as the Calzetti models does not include a contribution from spatially extended, non-attenuated ionized gas. The M31 data on right panel are color-coded according to their [SII]/H$ \alpha $ ratio (yellow for DIG and blue for HII region dominated) .}
\label{fig:FigDiff}
\end{figure*}

\clearpage

In conclusion, unlike the previously used Calzetti models of dust/gas distribution (combining HII regions, stars and dust), our results indicate that by adding a weakly- or unattenuated DIG component to those models, the attenuation can change significantly. The contribution of a DIG component also  yields higher [SII]/H$\alpha $ ratios than in HII regions, which can not be explained by the Calzetti models alone (see Section 4.3). Therefore the  contribution  of the DIG component and its relative vertical distribution (i.e. scale-height)  compared to the dust disk can explain the discrepancy between the nearby galaxies analyzed by K13 and our result in  M31.

\subsection{[SII]$(\lambda6717+\lambda6731)$/H$\alpha $ vs. A$ _{\rm V} $ diagram}
\label{Sec: 4.3 SIIHa vs Av}

   In the case where the DIG presides outside the dust disk and does affect our observations, we expect areas with higher [SII]/H$\alpha$ and the same $\Sigma_{\rm dust}$ to have a lower $A_{\rm V}$. There would be no trend between A$ _{\rm V} $ and [SII]/H$\alpha$ (at the same  $\Sigma_{\rm dust}$) if both components lie within the dust disk.
  
   In Fig. \ref{fig:FigDiff} our M31 data are color-coded according to their [SII]/H$\alpha $ ratio (with 0.35 as a threshold). Data points with lower ratio indicate areas with HII regions, while those with higher ratios mark DIG-dominated areas. Our M31 fields do not show a clear trend between the [SII]/H$\alpha$ ratio and $A_{\rm V}$, with Field 4 even exhibiting lower line ratios for regions lying further away from the foreground screen model. Due to possible changes of the intrinsic Balmer lines ratio in DIG dominated spaxels,  A$ _{\rm V} $ values of those spaxels could drop by down to 0.3-0.5 mag from those currently shown in Fig. \ref{fig:FigDiff}. K13 showed that there is a trend for a higher [SII]/H$ \alpha $ ratio at lower attenuation in all their galaxies, which they attribute to the fact that HII regions are located within dusty birth clouds (Fig. 4 in K13).  As the two lines are very close in wavelength,  extinction alone can not account for the change in the [SII]/ H$\alpha$ line ratio  (at A$ _{\rm V}$=3 mag the change is only 0.02 in the ratio).  
 
      Fig. \ref{fig:FigEnv}  shows [SII]/H$\alpha $ as a function of A$ _{\rm V} $ for regions within nearby galaxies (left panel) and M31 (right panel), color-coded by the dust column density. On both panels, the contours indicate the   distribution of the data for nearby galaxies at 2.7$ '' $ resolution.  On the left panel of Fig. \ref{fig:FigEnv}, the regions within nearby galaxies are convolved to 18$ '' $ resolution (spanning physical scales between $ \approx $300 and $ \approx $2000\,pc). NGC 2146 (X symbols)  is an outlier in this diagram as it has a high inclination, high dust column galaxy, with clear shock driven outflows (\citealt{Kreckel14}). On the right panel of Fig. \ref{fig:FigEnv}, our M31 data are shown at  24.9$ '' $ resolution ($ \approx $100\,pc) and the integrated fields.

      In general, the change in spatial scales affects our line ratios by diluting the flux from  compact, resolved objects like HII regions.   HII regions within M31 (with [SII]/H$\alpha $ $<$ 0.3 at 10pc scales) exhibit higher [SII]/H$\alpha $ ratios when convolved to 100pc scales. The same diluting effect causes a decrease in the [SII]/H$\alpha $ ratio for regions dominated by extended DIG emission.  
      
      Some areas (20\% of all regions and almost half of Fields 2, 4 and 5) are consistent with shock excitation ([SII]/H$\alpha$ $>$ 0.5, \citealt{Kewley06}), however they also show large uncertainties (due to low S/N as they have low surface brightness).  

      The left panel in Fig. \ref{fig:FigEnv} does not show a clear trend for lower attenuation at higher [SII]/H$\alpha$ ratios. However some of the individual galaxies in the K13 sample (like  NGC 3627, NGC 6946 and NGC 3077) do show a weak trend for increased [SII]/H$\alpha$ towards lower A$ _{\rm V} $. Shown in the right panel of Fig. \ref{fig:FigEnv}, the M31 data actually suggests an increase of the [SII]/H$\alpha$ ratio with A$_{\rm V}$. However there is a large scatter, and this trend is predominantly driven by variation between the fields, suggesting some dependence on location and physical characteristics of ISM (see section 4.4).  
      
   We could miss certain spaxels with both higher [SII]/H$\alpha$ ratio and A$ _{\rm V} $ due to faintness of the lines (caused by both low surface brightness and high attenuation). Such spaxels could possibly  change the trends seen in Fig. \ref{fig:FigEnv}.

 There is a possibility that the gas-phase  metallicity has an effect on the global trend in line ratios. Like most galaxies, there exists a radial metallicity gradient in M31 meaning the metallicity increases from Fields 1 to 5 (Table 3 in  \citealt{Kapala15}; \citealt{Zurita12, Draine14}). This  higher metallicity is associated with  the higher DGR and the higher  [SII]/Ha ratio. However, this ratio is affected not only  by metallicity, but also by ionization parameter and temperature of the gas, and this degeneracy causes issues in the analysis of the ratio and  A$ _{\rm V} $.

Besides the difference in trends,  thresholds and biases on A$ _{\rm V} $, M31 shows higher attenuation at fixed dust mass surface density, and higher [SII]/H$ \alpha $ ratio at fixed A$ _{\rm V} $ (especially at higher A$ _{\rm V} $) compared to the other nearby galaxies (as seen on Fig. \ref{fig:FigMain} \& \ref{fig:FigEnv}). More complicated models are required to better understand the affects of the different relative scale heights of dust, HII regions and DIG on the line ratios and on A$ _{\rm V} $.

   \subsection{Modeling the impact of geometry}
   \label{Sec: 4.3 Toys}

In this sub-section we investigate two simple models that examine how the vertical distribution (scale heights) and intensity of the HII and DIG phases compared to the dust layer can change the observed values and relations between $\Sigma_{\rm dust}$, A$_{\rm V}$, and the [SII]/H$\alpha$ ratio.  
   
   The first simple model (Fig. \ref{fig:FigToy1}) represents a face-on  galactic disk composed of only a thin layer of HII regions (with [SII]/H$ \alpha=0.15 $), a uniform dust screen, and a DIG component in front (with [SII]/H$ \alpha=0.75 $), all along one line of sight. We took those specific SII/H$ \alpha $ ratio for simplicity, but in a real ISM those values could change.  In this model, one resolution element includes emission from an HII region, that suffers an attenuation A$ _{\rm V}^{real} $ from the dust screen. The emission from the DIG is assumed to be unattenuated. The variables that are changed in the model are: 1) the intrinsic attenuation A$ _{\rm V}^{real} $ (directly proportional to $ \Sigma_{\rm dust} $) and 2) the ratio of intrinsic intensities from the DIG and the HII region (parameter X=I$ ^{DIG} $/I$ ^{HII} $). We emphasize that I$ ^{HII} $  is not attenuated and can not be associated with the value F$ ^{ext} $ described in the previous section. By varying these parameters, we can then explore the changes in the observed [SII]/H$ \alpha $ ratio (hereafter labelled S2) and A$ _{\rm V}^{\rm obs} $. This toy model is insensitive to spatial scales and to the amount of flux, as all variables are relative to each other. 
  
\begin{figure*}[ht!]
\centering
\includegraphics[width=0.95\linewidth]{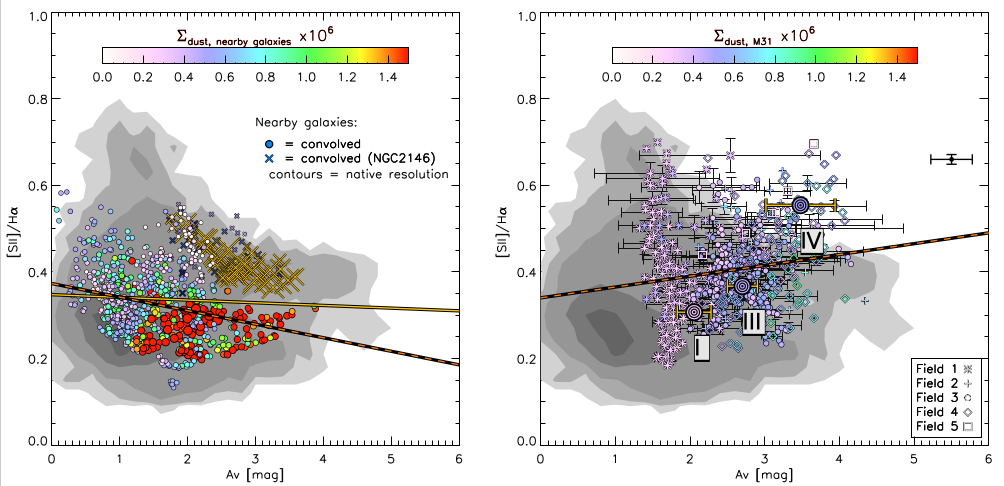}
 \caption{[SII]($ \lambda $6716+$ \lambda $6730)/H$ \alpha $ vs. A$ _{\rm V} $ diagrams of nearby galaxies (\textit{left}) and M31 (\textit{right}). Contours on both diagrams represent the K13 data points at native PPaK (24.9$ '' $) resolution. \textit{Left-} Convolved (18$ '' $ resolution) data points for the K13 nearby galaxies, coloured and in sizes according to $ \Sigma_{\rm dust} $. Yellow line shows trend of convolved data of all galaxies while orange dashed line shows the trend of the sample without NGC 2146. \textit{Right-} Convolved (24.9$ '' $ resolution) and integrated field data points for M31, also coloured by $ \Sigma_{\rm dust} $ values. Orange dashed line shows trends of the convolved data. Error bars are shown for a representative sample of the data, and the median value of all uncertainties (right corner) is also shown.  }
\label{fig:FigEnv}
\end{figure*}

\begin{figure*}[ht!]
\centering
\includegraphics[width=0.95\linewidth]{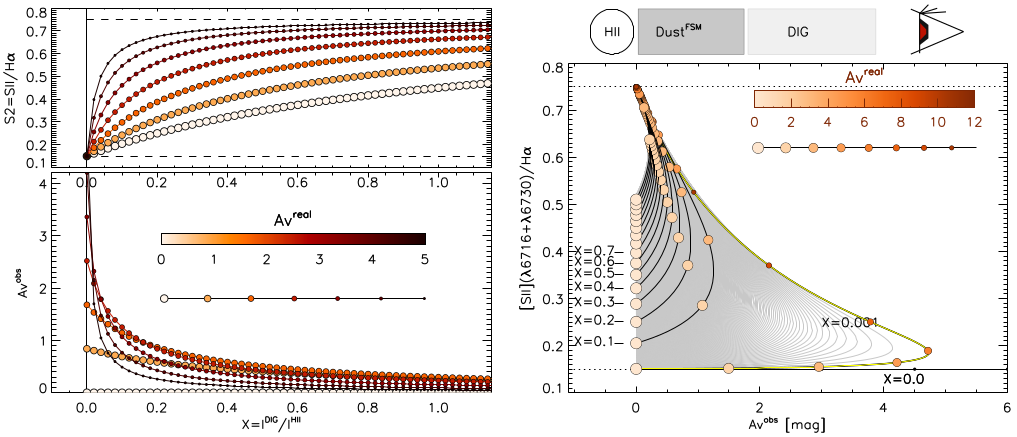} 
\caption{ A simplified model (diagram presented above the right panel) that explores the relationship between the observed line ratios and the spatial geometry of dust, HII and DIG regions in LOS of a face-on disk. In this model, between the observer  and the HII region (with an intrinsic brightness I$ ^{HII} $ and S2=[SII]/H$ \alpha $ ratio of 0.15) is a layer of dust  screen (labeled as Dust$ ^{\rm FSC} $ and proportional to A$ _{\rm V}^{\rm real} $) and DIG (with an intrinsic brightness I$ ^{DIG}=X\cdot $I$ ^{HII} $ and S2=0.75). \textit{Left-} As an input to the model, we have the amount of dust (A$ _{\rm V}^{real} $), upper and lower limits of S2 (0.15 and 0.75) and ratio between HII and  DIG intensity (or X). The correlation between those parameters (upper left panel) and the resulting observed attenuation A$ _{\rm V}^{obs} $ can be seen in the lower left panel. Notice that the observed attenuation does not follow the real attenuation and that there are overlaps where one value of A$ _{\rm V}^{obs} $ can have multiple values of $ \Sigma_{\rm dust} $ (depending on X). \textit{Right-} The correlation between the observed S2=[SII]/H$ \alpha $ ratio and the observed attenuation A$ _{\rm V}^{obs} $ with different ratios of DIG and HII intensities (shown as X) and different values of $ \Sigma_{\rm dust} $ (shown through A$ _{\rm V}^{real} $). Notice the resemblance between this image and those on the left panel of Fig. \ref{fig:FigEnv}. See text for more details. }
\label{fig:FigToy1}
\end{figure*}

  On the left panel of Fig. \ref{fig:FigToy1}, we show the correlation between  S2, the intrinsic and observed attenuation, and the ratio of intrinsic intensities from the DIG compared to from HII regions (parameter X).   Our model indicates that different combinations of emission from DIG and HII regions can generate the same observed S2 and A$ _{\rm V} $ values for a range of dust masses (i.e. $ \sim $A$_{\rm V}^{\rm real}$).   This is because of the large impact the DIG emission has on the S2 and A$ _{\rm V} $ values when the flux from the HII region is highly attenuated.  Combining that effect with the dependence of S2 on the dust and the DIG, our model resembles the behavior observed for nearby galaxies (Fig. \ref{fig:FigEnv}). However this model can not explain the data for M31, where we observe high values of [SII]/H$ \alpha $ and  A$ _{\rm V} $, nor explain the offset in attenuation between regions at fixed dust mass surface density in M31 and the other nearby galaxies.
  
\begin{figure*}[t]
  \centering
  \includegraphics[width=0.9\linewidth]{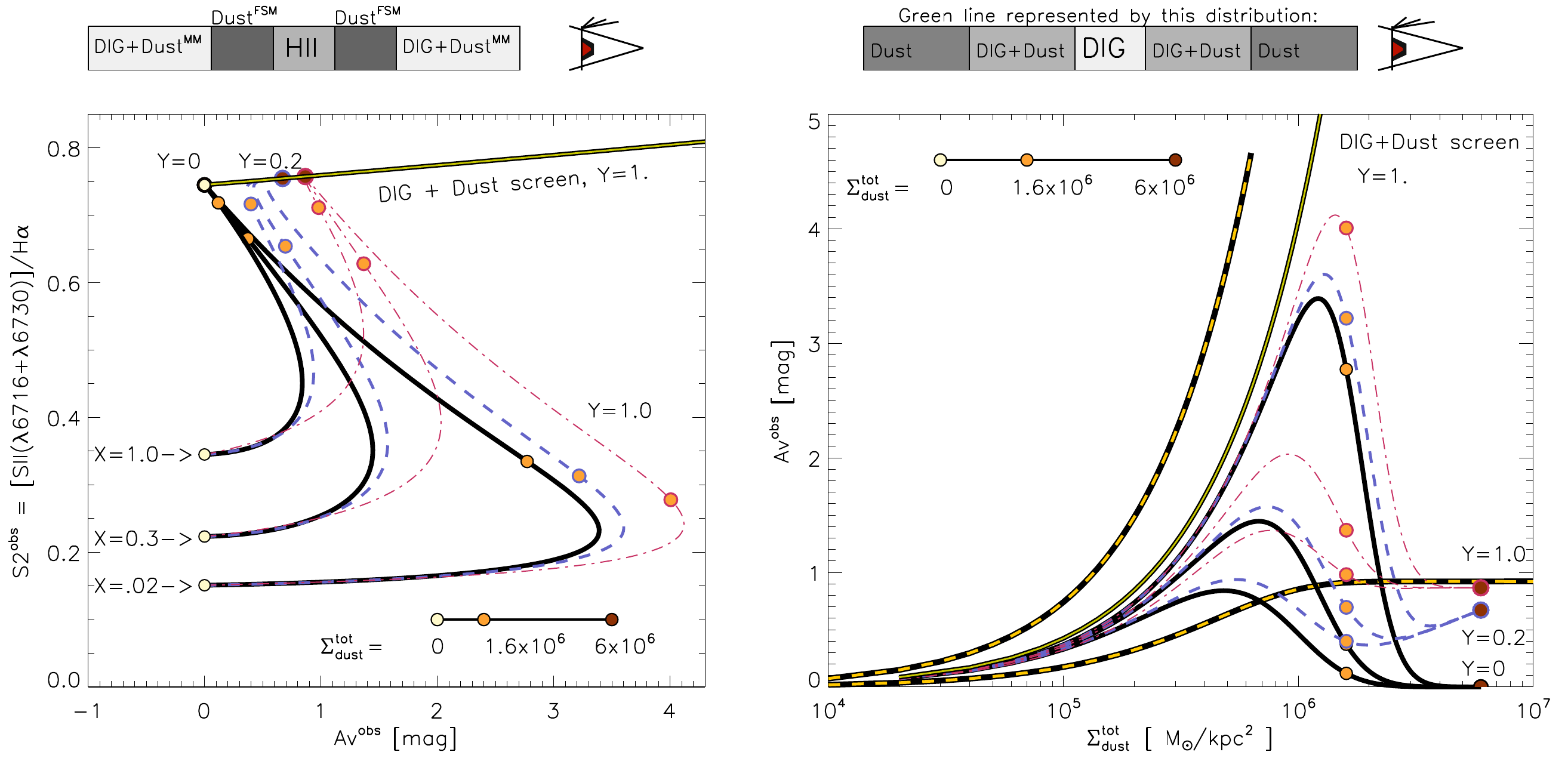}
  \caption{A second model (left panel on top), similar to the previous model (Fig. \ref{fig:FigEnv}), but with a  more realistic ISM distribution: DIG (mixed with the diffuse dust layer) and the dust screens lie on both sides of the HII region. The left panel shows [SII]/H$ \alpha $ vs. A$ _{\rm V}^{obs} $, while the right panel shows A$ _{\rm V}^{obs} $ vs. $\Sigma_{\rm dust}^{\rm tot} $. The diagrams show the relationship between the observed attenuation (A$ _{\rm V}^{obs} $),  the [SII]/H$ \alpha $ ratio (S2) with the model variables: the observed total dust column density ($  \Sigma_{\rm dust}^{\rm tot} $), the relative intensities of the DIG and HII regions (X=I$_{\rm DIG}$/I$_{\rm HII}$),  and the ratio of the  dust mixed with DIG to the total dust column (Y=$\Sigma_{\rm dust}^{DIG} $/$\Sigma_{\rm dust}^{tot} $). We assumed the same specific intrinsic S2 values for the gas in HII and DIG as in Fig. \ref{fig:FigToy1}. Values of Y=0, 0.2 and 1.0 are shown in black (full), blue (dashed) and red (dot-dashed) lines.  We also include the  green, full line that represents a scenario   (shown above the right panel) where the 90\% of DIG is mixed with dust  (while 10\% lie in the center and is not mixed with dust). This scenario also assumes length of the dust layer twice as large  as the length of the ionized gas layer, thus representing our assumption about dust/gas distribution of M31. With this scenario we can reproduce the regions with higher A$ _{\rm V}^{obs} $ and SII/H$ \alpha $ ratios seen in Fig. \ref{fig:FigEnv}. }  
  \label{fig:FigToy2}
  \end{figure*}

\begin{figure}[h!]
  \centering
  \includegraphics[width=1.0\linewidth]{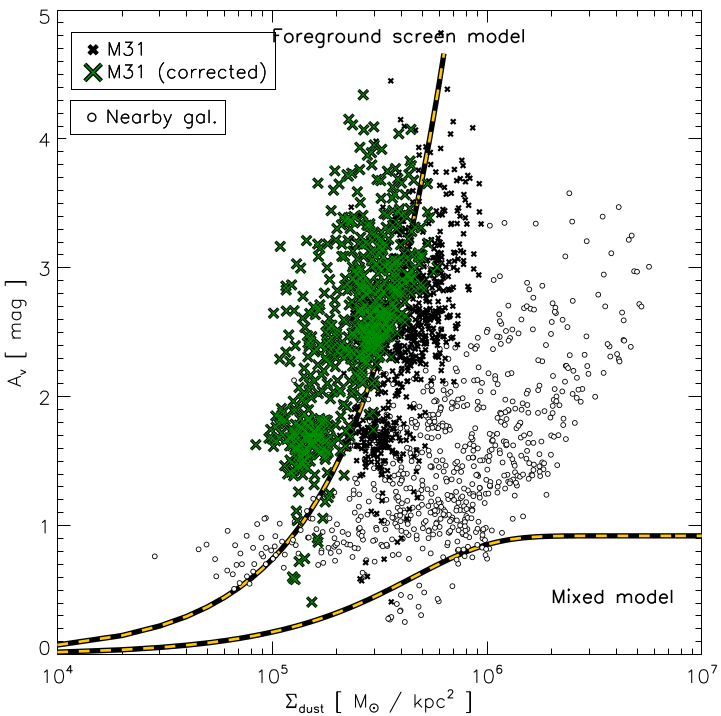}
  \caption{ A$ _{\rm V} $ vs. $ \Sigma _{\rm dust} $ before and after correcting dust mass values using the empirically derived renormalization formula 9 in \citet{Planck16b}. Crosses indicate the data for M31 and circles the data for the nearby galaxies.  }
  \label{fig:AppendixA}
  \end{figure}

  The second model (Fig. \ref{fig:FigToy2})  has a more realistic ISM distribution where the layers of  DIG and diffuse dust lie on both sides of the HII region and are mixed, as observed by \citet{Howk12}. The left panel shows [SII]/H$ \alpha $ vs. A$ _{\rm V}^{obs} $, while the right panel shows A$ _{\rm V}^{obs} $ vs. $\Sigma_{\rm dust}^{\rm tot} $. In this model  we assume that a certain amount of diffuse dust is mixed with the DIG. The mass of mixed dust is given by the ratio of mixed to total dust mass (parameter Y=$\Sigma_{\rm dust}^{\rm DIG} $/$\Sigma_{\rm dust}^{tot} $). $\Sigma_{\rm dust}^{tot} $ is equivalent to the total observed dust and it is a sum of the dust layer around HII region which acts as a screen  (labeled as Dust$ ^{\rm FSC} $ in Fig. \ref{fig:FigToy2}) and the dust layer mixed with DIG (labeled as Dust$ ^{\rm MM} $ in Fig. \ref{fig:FigToy2}).

\clearpage

  On the second model, shown in Fig. \ref{fig:FigToy2}, a case of Y=0 represents the case where the dust is not mixed with DIG and there is no Dust$ ^{\rm MM} $. That is the case similar to the first model shown in Fig. \ref{fig:FigToy1}. When we redistribute the total dust so that some part of it is mixed  with the DIG (thus rising value Y), the dust dims the DIG emission and increases the measured  attenuation. We present an extra scenario in Fig. \ref{fig:FigToy2} where  there is only DIG and dust and the scale height of the DIG is smaller than the scale height of the dust layer. That scenario shows higher A$ _{\rm V}^{obs} $ and higher SII/H$ \alpha $ ratios compared to previous cases.  With that scenario, we can reproduce the regions with higher attenuation and higher [SII]/H$ \alpha $ ratio noticed in Fig. \ref{fig:FigDiff} and \ref{fig:FigEnv}.

    Comparing the K13 and M31 data seen in  Fig. \ref{fig:FigMain} and \ref{fig:FigEnv} with the models in   Fig. \ref{fig:FigToy1} and \ref{fig:FigToy2}, we notice a complex behaviors and the strong effect the DIG vertical distribution has on the observed A$ _{\rm V}^{obs} $ and  SII/H$ \alpha $ ratios. 
    
   In the nearby galaxies studied by K13, the DIG components could be more extended and not so well mixed with the dust. We draw this conclusion by comparing  the observed ratios in Fig. \ref{fig:FigEnv} with the simple models on Fig. \ref{fig:FigToy1} and \ref{fig:FigToy2}. The likely  difference between those galaxies and M31 is the intensity and relative vertical distribution of dust and DIG. 
 
 Unlike K13, the higher spatial resolution  data in M31 enables us to clearly distinguish between compact HII regions and those regions with dust and DIG only. Our results indicate that the DIG component in M31 could be smaller in intensity and scale height, and be well mixed with the dust layers in the disk.  An additional layer of diffuse dust, extending to even larger scale heights than the DIG, would increase the attenuation even more.

    The high inclination of M31 likely plays an additional role in increasing the attenuation. Nevertheless, K13 did not report any significant correlation or change in their relation with inclination. For example, the highly inclined galaxy NGC 3627 (similar to M31) shows lower Av than M31, while NGC 4321 has higher Av values despite its lower inclination (30$ ^{\circ} $) and similar $\Sigma_{\rm dust} $ as NGC 3627. NGC 2146 and NGC 7331 are both highly inclined (60$ ^{\circ} $ and 70$ ^{\circ} $ as M31) but have slightly lower A$ _{\rm V} $ value than those seen in M31.

    \subsection{Effects of location on the relative vertical distribution of dust and DIG}
    
 \label{Sec: 4.4 Env}

 Given these findings, the question that arises is: why in M31 do we observe a different mixture and scale heights between all three components (DIG, HII regions and the dust) compared to other galaxies? To explain this, we consider the effects of probing different locations within the galactic disk. 
    
   We note that the locations and radii of the fields observed in M31 and in  K13 are different. In M31, we observe five small fields within the spiral arms of M31, where the most central field is still  $ \sim $6\,kpc from the galaxy center (equal to 0.28 R$ _{25} $; Tab. \ref{tab:Tab01}, \citealt{Zurita12, deVaucouleurs91}).  The regions observed by K13 are more central ($<$ 0.2 R$ _{25} $, except for one galaxy with $<$ 0.4 R$ _{25} $ ), and composed of  bulges, spiral arms and inter-arm regions. The global SFR and $ \Sigma_{\rm SFR} $ for the nearby galaxies studied by K13  are much higher than that of M31 (${\rm SFR} \sim 1$ M$ _{\odot} $\,yr$ ^{-1} $; \citealt{Williams03}) and the observed fields ($ \Sigma_{\rm SFR} \sim 0.01$ M$ _{\odot} $\,yr$ ^{-1} $\,kpc$ ^{-2} $; see Fig. \ref{fig:AppendixB} in Appendix B). 
   
   If we assume that the dust follows the gas, as seen in  \citet{HughesBaes14} and \citet{Holwerda12} with some dependence on dust temperature, then we can argue that the scale height of the dust is equal to the gas scale height and follows it. In several observations  the scale heights of the  gas has been found to increase with galactic radius (\citealt{Sancisi79}, \citealt{Braun91}, \citealt{Rupen91}, \citealt{Scoville93}, \citealt{Olling96},  \citealt{Garcia99}, \citealt{Wong02}, \citealt{MatthewsWood03}, \citealt{Obrien10}, \citealt{Velusamy14}, \citealt{Yim14},  \citealt{Zschaechner15}). The increase of the gas scale height can be understood via the  gravitational and hydrodynamical equilibrium of the gas, and its dependence on gas vertical velocity dispersion, stellar volume density and gas surface density (\citealt{Koyama09}, \citealt{Pety13}). If the dust and  molecular gas are correlated, then this mean that the dust scale height is rising with galactic radius.    
   
   On the other hand, the DIG scale height and intensity are correlated with the number and brightness of HII regions and star formation activity (\citealt{Dettmar90}, \citealt{Rand96}, \citealt{Wang99}, \citealt{Collins00}, \citealt{Oey07}). Numerous, brighter star-forming regions that are more energetically active (\citealt{Tyler04}, \citealt{Yasui15}) can affect the production, destruction and distribution of dust and also the ionization of the DIG. This can result in a more prominent and extended DIG component.  Moreover, \citet{Dettmar90} noticed a decrease of the H$ \alpha $ scale height with galactic radius (with the highest scale height seen in the starbust regions in the center). We argue that if the scale height of the DIG is proportional to the star formation intensity, then the decrease of star formation with radius would indicate a decrease of the scale height of DIG component. 
\citet{Regan06} and \citet{Bigiel10} found the FUV and 8$ \,\mu $m emission (both correlated to star formation) are decreasing with galactic radius and have their highest values  in the centers of galaxies. Furthermore, \citet{Garcia99} noticed the existence of H$ \alpha $ chimneys spanning the extra-planar area from the galactic center with scale height larger than that of the CO.

Since the dust scale height  is likely to increase with galactic radius while the DIG scale height should decrease with galactic radius, differences in the relative DIG/dust geometry and scale heights of galaxy centers (as in K13 galaxies) and galactic outskirts (as in M31) may explain the differences in our derived attenuation.

\section{Wider implications}

 In order to determine the spatial distribution of dust compared to the ionized gas, we use the models derived by \citet{Caplan86}. The only components of that model are the dust and ionizing stars  and their  spatial configuration that affects the observed attenuation (\citealt{Caplan86, Calzetti94}).  Our results (Fig. \ref{fig:FigMain}) show that the regions we observe in M31 follow closely a foreground screen model, while K13 showed that most of the regions they selected in a sample of nearby galaxies fall between the mixed and screen models. Furthermore, \citet{Liu13} in M83 find that the data points of the central region follow more a mixed model, while the outer regions are in agreement with a foreground screen model. Tests show that neither inclination (K13) nor changes in spatial scales (Fig. \ref{fig:FigScales}) can explain the differences between those results. However, by degrading the spatial resolution of the data (toward $ \geq $100-200\,pc scales) in the central region of M83, \citet{Liu13} showed that their data begin to follow more closely a foreground screen model. While such results are suggestive, their interpretation should be done in caution, given the large uncertainties, large scatter in the data and the small separation of the models in the diagrams.

 Our observation of varying [SII]/H$ \alpha $ line ratios and modeling of  the impact of the relative vertical distribution of dust and DIG  shows that the DIG component needs to be accounted for in  the models of \citet{Caplan86}, \citet{Calzetti94} and \citet{Wong02}.  From our observations (Fig. \ref{fig:FigMain}, \ref{fig:FigDiff} and \ref{fig:FigEnv}) and simple models  (Fig. \ref{fig:FigToy1} and \ref{fig:FigToy2}), we conclude that the differing scale heights of dust and gas in galaxies plays a large role in the measured extinction.
 
 In general, we argue that the star-forming  regions are born in a cocoon of dust, with the formed massive stars clearing the dust over time (\citealt{Dreher81, Verley10}). The varying scale heights of dust and ionized gas (HII regions and DIG) with radius in galaxies can change the measured attenuation.  At large galactocentric radii, especially in M31 with its low star formation rate, the HII regions remain embedded in their dusty cocoons that act as foreground screens. The DIG component is not as prominent in intensity or scale height. The DIG in M31 is well mixed with the thin dusty disk and additionally attenuated by the diffuse dust layer that acts as a screen. The high inclination of M31 also contributes to the higher attenuation.

 On the other hand, in the more central and more active regions of galaxies (as in K13),  the dust may be blown away by the numerous bright star-forming regions. The DIG will also be more prominent and extended, with a scale height larger than the scale height of dust. This leads to the DIG being less attenuated by dust and causing it to have a larger effect on observations of attenuation.

\section{Summary}

  Using Integral Field Spectroscopy (IFS) and IR photometry in five 680\,pc $ \times $ 900\,pc fields in M31, we explore the relative spatial distribution of dust and ionized gas. This is done at $ \sim $100 pc scale resolution by  comparing the attenuation  (A$ _{\rm V} $) determined from the Balmer decrement and the dust mass surface density determined from fitting to the IR SED photometry ($\Sigma_{\rm dust}$).   We compare the results with two widely used theoretical models of the dust distribution (mixed and screen models, \citet{Calzetti94}) and with previous results from the literature for eight nearby galaxies (\citet{Kreckel13}, K13).

  Our results show that the dust is approximately distributed as an uniform screen around the ionized gas in M31 (Fig. \ref{fig:FigMain}). This is distinct from the galaxies observed by K13 , which show attenuations between those predicted by the mixture of the foreground screen model and the mixed model (where the dust is uniformly mixed with the gas). 

  Variations in the spatial resolution do not appear to explain the differences in the measured attenuations (Fig. \ref{fig:FigScales}). 
  
The contribution from a small amount of non-attenuated gas emission in front of the dust disk can significantly lower the observed attenuation (Fig. \ref{fig:FigDiff}). This can lead to biases in observations of nearby galaxies. This additional gas can be associated to the diffuse ionized gas (DIG).

 We also analyze the observed [SII]/H$\alpha $ ratio and its correlation with A$ _{\rm V} $ (Fig. \ref{fig:FigEnv}).  Although our data do not show a clear trend, we found a difference in behavior between M31 and the nearby galaxies in K13. The M31 data show a slight increase of the ratio with A$ _{\rm V} $, while the K13 galaxies show the opposite trend. 
 
 Using two simple models, we explore the relationship between the observed line ratios and the relative spatial geometry of the dust, HII regions and DIG gas (Fig. \ref{fig:FigToy1} and \ref{fig:FigToy2}). These models indicate that the relative vertical distribution and contribution of the DIG and the dust  play a crucial role in changing the observed [SII]/H$\alpha $  and  A$ _{\rm V} $ values at given values of the total dust column.
 
  The difference in the results of M31 and K13 can be explained by the fact that the  M31 fields lie at large galactocentric radii, whereas the K13 span fields which are in the centers of the galaxies. These differences in radii are associated with differences in the scale heights of the dust, DIG and HII regions which impact the measured attenuation.

\begin{acknowledgments}

Tomi\u{c}i\'{c} N. and Kreckel K. acknowledge grants SCHI 536/8-2 and KR 4598/1-2 from the DFG Priority Program 1573. B.G. gratefully acknowledges the support of the Australian Research Council as the recipient of a Future Fellowship (FT140101202). G.B. is supported by CONICYT/FONDECYT, Programa de Iniciaci\'{o}n, Folio 11150220.  

We thank the anonymous referee for useful comments that helped to improve the paper. Furthermore, we thank to I-Ting Ho (MPIA) for useful feedback and discussions which also helped to improve the paper.

This work is based on observations collected at the Centro Astron\`{o}mico
Hispano Alem\`{a}n (CAHA), operated jointly by the Max-Planck Institut f\"{u}r
Astronomie and the Instituto de Astrofisica de Andalucia (CSIC), and is
also based on observations made with Herschel. Herschel is an ESA space
observatory with science instruments provided by European-led Principal
Investigator consortia and with important participation from NASA. PACS
has been developed by a consortium of institutes led by MPE (Germany)
and including UVIE (Austria); KU Leuven, CSL, IMEC (Belgium); CEA, LAM
(France); MPIA (Germany); INAF-IFSI/OAA/OAP/OAT, LENS, SISSA (Italy);
and IAC (Spain). This development has been supported by the funding
agencies BMVIT (Austria), ESA-PRODEX (Belgium), CEA/CNES (France), DLR
(Germany), ASI/INAF (Italy), and CICYT/MCYT (Spain). SPIRE has been
developed by a consortium of institutes led by Cardiff University (UK)
and including Univ. Lethbridge (Canada); NAOC (China); CEA, LAM
(France); IFSI, Univ. Padua (Italy); IAC (Spain); Stockholm Observatory
(Sweden); Imperial College London, RAL, UCL-MSSL, UKATC, Univ. Sussex
(UK); and Caltech, JPL, NHSC, Univ. Colorado (USA). This development has
been supported by national funding agencies: CSA (Canada); NAOC (China);
CEA, CNES, CNRS (France); ASI (Italy); MCINN (Spain); SNSB (Sweden);
STFC (UK); and NASA (USA).

\end{acknowledgments}

\bibliographystyle{aasjournal}
\bibliography{NT_Article_Attenuation_M31}

 \begin{figure}[h!]
  \centering
  \includegraphics[width=1.0\linewidth]{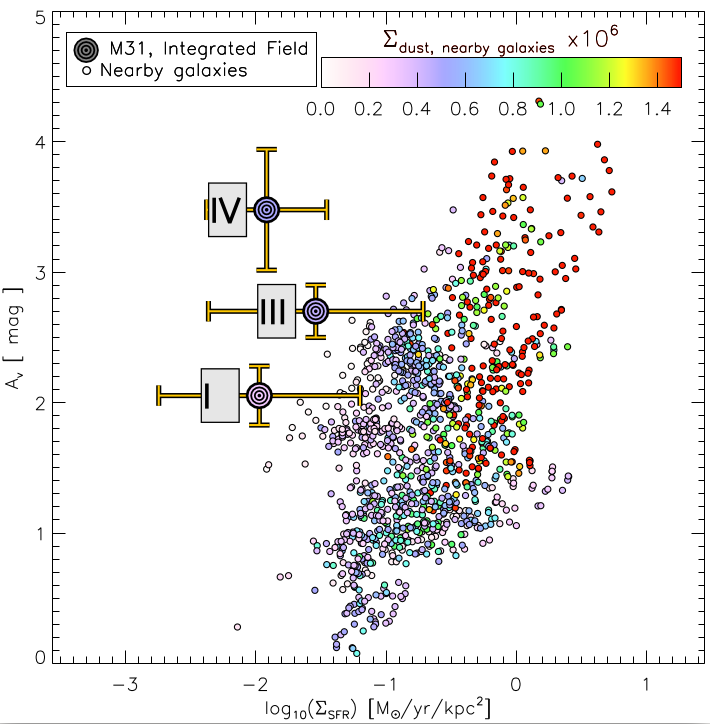}
  \caption{  A$ _{\rm V} $ vs. $ \Sigma_{\rm SFR} $ (from extinction corrected H$ \alpha $) of M31 (big circles) and K13 galaxies (small circles). Colors corresponds to  $ \Sigma _{\rm dust} $ of the data. M31 data are at 0.6\,kpc$ \times $0.9\,kpc resolution, while the K13 are at 0.3\,kpc-2\,kpc resolution.    }
  \label{fig:AppendixB}
  \end{figure} 

\section{Appendix A} \label{Sec: Appendix A}

 The dust mass surface density maps used in this work were derived by the SED fitting method explained in \citet{Draine14} and \citet{M31Groves12}, with an assumption of the \citet{DraineLi07} dust  model (here labeled as DL07). For details see Sec. \ref{Sec: Data}. 

  Two papers (\citealt{Dalcanton15}, \citealt{Planck16b}) recently tested the validity of the  DL07 model by measuring the dust column density via extinction of the light from various background sources.  \citet{Dalcanton15} used the old stellar population in the thick disk of M31 for the light sources, while \citet{Planck16b} used quasi-stellar objects (QSO) optical photometry.  Both studies indicate a discrepancy by the factor $ \sim $2.5  between the DL07 estimates for A$ _{\rm V} $ and  A$ _{\rm V} $ estimated from the background sources. Furthermore, comparison with independent far-IR observations (Planck, Herschel, Wise and IRAS) have shown that this offset is not due to uncertainties in the Herschel photometry \citep{Verstappen13, Planck16b}.

 \citet{Planck16b} propose an empirical renormalization of the dust mass derived from DL07 as a function of the DL07 ionization parameter U$ _{\rm min} $ (see Formula 9 in \citealt{Planck16b}). Fig. \ref{fig:Appendix} shows the A$ _{\rm V} $ vs. $ \Sigma _{\rm dust} $ diagram for M31 and the nearby galaxies from K13, before and after using renormalization proposed by \citet{Planck16b}.  The regions observed in K13 have U$ _{\rm min} $  $>$ 1.  As the renormalization is not calibrated in this high U$ _{\rm min} $ regime, we do not include here any renormalization of the K13 regions.   However, extrapolating the renormalization would make $ \Sigma _{\rm dust} $ of the K13 galaxies larger, thus widening the relative disagreement between M31 and K13.

  The renormalized $ \Sigma _{\rm dust} $ values for M31 are lower than before,  pushing the values above the foreground screen limit lines. Given our confidence in our spectrophotometric calibration, we can only explain these relatively extreme Av values as a result of either different physical dust properties or a non-uniform dust distribution (i.e. selective extinction of the HII regions). Such differences could lead to variations in extinction laws, DGR and R$ _{\rm V} $.

\section{Appendix B} \label{Sec: Appendix B}

 We investigate how the attenuation is affected by the different spatial resolutions between our M31 data and the nearby galaxy sample of K13.  In Fig. \ref{fig:AppendixB}, we present A$ _{\rm V} $ vs. star-formation rate density ($ \Sigma_{\rm SFR} $; from extinction corrected H$ \alpha $) for both M31 and the K13 galaxies. The data points are color-coded base on $ \Sigma_{\rm dust} $. The K13 galaxies are at 0.3\,kpc-2\,kpc resolutions. We match our M31 data to the resolutions of K13 by integrating the fluxes in 0.6\,kpc$ \times $0.9\,kpc fields. The lower resolution M31 data show significantly lower $ \Sigma_{\rm SFR} $ than the K13 galaxies, making a direct comparison between M31 and the K13 galaxies at a fixed $ \Sigma_{\rm SFR} $ impossible.

\end{document}